\documentclass[double]{article}
\usepackage{times}
\usepackage{algorithm}
\usepackage{algorithmic}
\usepackage{wrapfig}
\usepackage{verbatim}
\usepackage{amsmath}
\usepackage{graphicx}
\usepackage{subcaption}
\usepackage[active]{srcltx}
\usepackage{color}
\usepackage{lineno}
\usepackage{dirtytalk}
\usepackage{amsthm}
\usepackage{authblk}
\usepackage{listings}
\usepackage{tikz}
\usepackage{longtable}
\usepackage{setspace}
\usepackage{tikz}
\usetikzlibrary{positioning, arrows}
\usetikzlibrary{shapes.geometric, arrows}

\usepackage{epsfig,graphicx,amsfonts}
\usepackage{amsmath,amsthm,amssymb}
\usepackage{xcolor}

\usepackage{adjustbox}
\usepackage{multirow}
\usepackage{setspace}
\usepackage{hyperref}
\usepackage{comment}

\long\def\comment #1\commentend{}

\begin{document}

\title{\Large Chronic Stress, Immune Suppression, and Cancer Occurrence: Unveiling the Connection using Survey Data and Predictive Models}

\author{Teddy Lazebnik$^{1,2,*}$ and Vered Aharonson$^{3,4}$\\ \(^1\) Department of Information Science,  University of Haifa, Haifa, Israel \\ \(^2\) Department of Computing, Jonkoping University, Jonkoping, Sweden \\ \(^3\) Department of Basic and Clinical Sciences, Medical School, University of Nicosia, Nicosia, CY\\ \(^4\) School of Electrican and Information Engineering, University of the Witwatersrand, Johannesburg, ZA\\ \(^*\) Corresponding author: teddy.lazebnik@ju.se \\}

\date{ }

\maketitle 

\begin{abstract}
\noindent
\normalsize
Chronic stress was implicated in cancer occurrence, but a direct causal connection has not been consistently established. Machine learning and causal modeling offer opportunities to explore complex causal interactions between psychological chronic stress and cancer occurrences. We developed predictive models employing variables from stress indicators, cancer history, and demographic data from self-reported surveys, unveiling the direct and immune suppression mitigated connection between chronic stress and cancer occurrence. The models were corroborated by traditional statistical methods. Our findings indicated significant causal correlations between stress frequency, stress level and perceived health impact, and cancer incidence. Although stress alone showed limited predictive power, integrating socio-demographic and familial cancer history data significantly enhanced model accuracy. These results highlight the multidimensional nature of cancer risk, with stress emerging as a notable factor alongside genetic predisposition. These findings strengthen the case for addressing chronic stress as a modifiable cancer risk factor, supporting its integration into personalized prevention strategies and public health interventions to reduce cancer incidence. \\\\


\end{abstract}

\maketitle \thispagestyle{empty}
\pagestyle{myheadings} \markboth{Draft:  \today}{Draft:  \today}
\setcounter{page}{1}

\onehalfspacing

\section{Introduction}
\label{sec:introduction}
Cancer remains a leading cause of mortality worldwide \cite{intro_1}. Its occurrence and development are influenced by a complex interplay of genetic \cite{intro_2}, environmental \cite{intro_3}, and physiological factors \cite{intro_4}. Psychological stress is increasingly being considered a factor in cancer risk \cite{intro_5_1,intro_1}, and in the inhibition of cancer treatment efficacy \cite{intro_11,Mohan2022PsychosocialSA}. Chronic stress can disrupt homeostatic processes, leading to sustained hormonal and inflammatory responses \cite{intro_6}. These processes weaken the immune system, which is vital for defending against tumor development \cite{intro_7}. 

While physiological studies and animal models elucidated insights on possible mechanisms by which chronic stress can induce cancer, epidemiological human studies portrayed inconsistent results on this association \cite{Butow2018DoesSI}. An empirical validation of the stress-cancer link in human populations remains underexplored. Namely, currently, there is no clear evidence that chronic stress causes cancer occurrence in either a direct or throughout immune suppression. 

Machine learning (ML) has emerged as a powerful tool for the analysis of complex, multi-dimensional health data that can uncover hidden patterns and be used to study the connections in the data and make accurate predictions \cite{intro_14}. ML models have been successfully applied to cancer risk prediction \cite{intro_15_1,intro_15_2}, psychological health assessment \cite{intro_16_1,intro_16_2}, and immune response modeling \cite{intro_17_1,intro_17_2}. Nevertheless, no study leveraged ML to quantitatively assess the relationship between chronic stress, immune suppression, and cancer occurrence. 

In this study, we apply ML methods to survey data to investigate the association of chronic stress, through immune suppression indicators, to cancer occurrence. By integrating stress perception measures, immune function indicators, self-reported cancer history, and family cancer history, we aim to develop predictive models that identify patterns linking stress to cancer risk. Our ML analysis is complemented by statistical methods and causal models to corroborate the findings and to examine the stress to cancer causality. Using these models, we investigate a possible direct or immune suppression mitigated connection between chronic stress and cancer occurrence. 

The reminder of the manuscript is organized as follows. Section \ref{sec:rw} reviews the known factors between chronic stress and cancer progression, stress-induced immune suppression, and immune suppression influence on cancer emergence. Section \ref{sec:methods} presents the survey conducted, machine learning modeling, and the statistical analysis conducted. Section \ref{sec:results} outlines the obtained results. Finally, section \ref{sec:discussion} discusses the obtained results in their socio-clinical context and suggests future work.   

\section{Related Work}
\label{sec:rw}
In this section, we examine the interrelationships between chronic stress, immune suppression, and cancer occurrence, addressing each pair of factors independently.

\subsection{Chronic stress and cancer occurrence}
The associations between severe life events, anxiety, depression, and insufficient social support with increased cancer occurrence have been extensively reviewed in the past two decades \cite{Antoni2006TheIO,Zhang2025BurdenAR}. Stress management strategies, including pharmacological, physical, social, and psychological approaches, could be instrumental in cancer prevention efforts \cite{Cui2020CancerAS}. Recent epidemiological evidence suggests that chronic psychological stress may be a risk factor for cancer, particularly for breast, colorectal, lung, prostate, and pancreatic cancers \cite{Kruk2019PsychologicalSA,Dai2020ChronicSP,Mohan2022PsychosocialSA,Chiriac2017PsychologicalSA, Bahri2019TheRB}. 
However, results from both epidemiological and clinical studies were inconsistent  \cite{Lempesis2023RoleOS, Heikkil2013WorkSA,Butow2018DoesSI}. Moreover, the studies reveal methodological challenges in assessing the impact of stress on cancer occurence may hamper the identification of this association \cite{CortsIbez2022TheVO}. For instance, they could not elicit the contribution of smoking, alcohol consumption, and poor diet, which may be stress-induced lifestyle behaviors, to cancer occurrence \cite{Mohan2022PsychosocialSA,Avgerinos2019ObesityAC}. 
The reliable measurement of chronic stress is another significant challenge of these population-based studies \cite{Pham2024ChronicSR}. 
Finally, the mechanisms by which chronic stress influences cancer are complex and not yet fully understood. Prolific research focused on elucidating the molecular mechanisms through which chronic stress facilitates tumor occurrence and progression, with immune suppression being identified as a key mechanism \cite{Alotiby2024ImmunologyOS}.

\subsection{Chronic Stress and immune suppression}
Chronic stress has been consistently linked to the suppression of immune function  \cite{Alotiby2024ImmunologyOS,Nakata2012PsychosocialJS,segerstrom2004psychological}, of both cellular and humoral immunity \cite{Dhabhar2013PsychologicalSA}. 
Recent research has highlighted mechanisms by which chronic psychological stress can modulate immune function. These include pathways involving the hypothalamic-pituitary-adrenal axis and the sympathetic nervous system \cite{intro_8_1,intro_8_2,Antoni2006TheIO,Alotiby2024ImmunologyOS}. Stress-related elevations in cortisol and catecholamines can suppress immune surveillance by reducing the activity of natural killer cells, impairing T-cell function, and altering cytokine profiles \cite{intro_9_1,intro_9_2,Gouin2008ImmuneDA, Hong2021ChronicSE}. Animal studies further demonstrated that chronic stress impairs T cell function through decreased T cell proliferation, altered cytokine secretion, and increased lymphocyte apoptosis \cite{ColonEchevarria2019NeuroendocrineRO, Kusnecov2002StressorinducedMO, Silverman2012GlucocorticoidRO}. Inducing low-grade chronic inflammation, this dysregulation increase proinflammatory factors, and suppresses the numbers, trafficking, and function of immunoprotective cells \cite{intro_13_1,intro_13_2}.
These changes may foster an environment conducive to tumor initiation and progression \cite{intro_10}. 
The autonomic nervous system, activated by stressors like anxiety, pain, sleep disorders, or depression, is considered to play a crucial role in this process \cite{intro_12_1,intro_12_2,intro_12_3}. 
The dysregulation impact on both innate and adaptive immunity was associated with attenuated vaccine responses, impaired control of latent viruses, and heightened inflammation in patients \cite{Antoni2006TheIO,Cohen2001PsychologicalSA}. Decreased activity of tumor-infiltrating T lymphocytes was associated with higher levels of stress in ovarian cancer patients \cite{Gouin2008ImmuneDA, MorenoSmith2010ImpactOS}, and with increased susceptibility to skin cancer \cite{Rolln2022CurrentKO}. 
 
The studies linking stress to immune suppression convey multiple limitations. A gap persists in the translation of laboratory findings to clinical applications. The individual differences in stress tolerance may limit the generalizability of results. The underlying mechanisms and directionality of associations between variables such as age, gender, socioeconomic status, and stress-induced immune suppression are not yet well understood. Understanding these mechanisms is crucial for developing strategies that could mitigate the adverse effects of stress on immune function and overall health. Finally, a direct link between these immune alterations and disease outcomes is often absent in studies. 

\subsection{Immune suppression and cancer occurrence}
Psychoneuroimmunology research highlighted the significant role of immune function in tumor behaviors \cite{doss2016changing,Zhang2020ChronicSI}. Immunosuppression, whether primary, secondary, or induced by cancer itself, is associated with increased cancer incidence, particularly of skin cancers, and often exhibits patterns observed during fetal development \cite{Liu2014TheRO,Troche2014SystemicGU, Kareva2020ImmuneSI}. Targeting tumor-induced immune suppression—encompassing regulatory cells, cytokines/chemokines, T cell exhaustion, and metabolic factors is a promising approach for improving cancer immunotherapy \cite{Stewart2011ImprovingCI, Coussens2013NeutralizingTC}. However, while evidence underscores that certain aggressive cancers tend to develop in immunocompromised hosts, the factors influencing this, including specific aspects of immune deficiency and variations in patient populations, remain underexplored \cite{Bajpai2020CancerIF}.

Taken jointly, the current literature suggests potential links between chronic stress, immune suppression, and cancer occurrence. The studies employ diverse methodologies for assessing stress and include various study populations, which can impact the generalisation of the findings. There remains a lack of sufficient data to conclusively determine the influence of gender, age, and socioeconomic status on these connections. Further research is necessary to explore how these relationships may differ across populations and to identify underlying mechanisms. Such research would benefit from the adoption of standardized stress measurement tools and the concurrent analysis of multiple moderating factors. Understanding the relationship between chronic stress, immune suppression, and cancer risk may allow for early identification of at-risk populations. Harnessing this understanding for the development of targeted prevention strategies that address stress-related modifiable risk factors may reduce cancer incidence \cite{Mohan2022PsychosocialSA}.

\section{Methods and Materials}
\label{sec:methods}
\begin{figure}[!ht]
    \centering
    \includegraphics[width=0.99\linewidth]{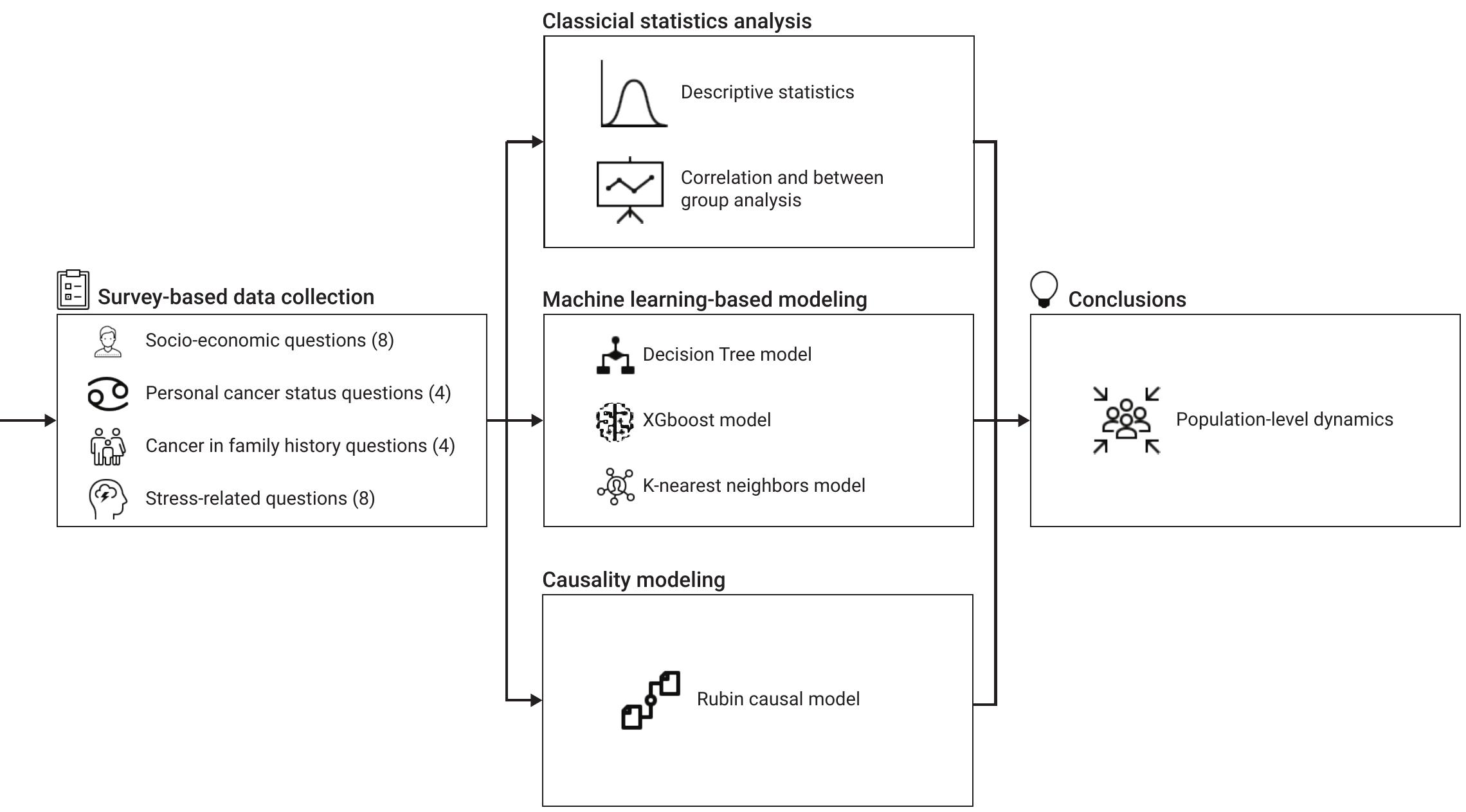}
    \caption{A schematic view of the methodological flow of this study.}
    \label{fig:flow}
\end{figure}

We employed a three-phased workflow, outlined in Fig. \ref{fig:flow}. Initially, we developed a four-section survey to gather data. Next, we analyzed the data using three complementary methods: descriptive statistics and linear regression model, non-linear machine learning-based regression models, and causal modeling. Based on the combined results, we interpreted potential population-level dynamics.

\subsection{Survey}
\label{sec:survey}
To gather real-world data on stress and cancer status, we conducted a questionnaire that consisted of four sections. The first section includes eight questions on the participant's socio-demographic and economic section (\(q^1_1-q^1_8\)): country, gender, age, marital status, number of children, education, income, and household size. These factors are used in two contexts: 1. a validation that the sample is broad and representative; and 2. an investigation of socio-economic factors' effect on the relation between stress and cancer occurrence. These factors are picked based on their documented importance for healthcare and social policy design \cite{qu_1,qu_2,qu_3,qu_4,qu_5}. The second section includes four questions on the participant's cancer status, focusing on personal experiences with the disease (\(q^2_1-q^2_4\)): if the participant had cancer and if so, what type, when, and whether the participant is still undergoing cancer treatment  \cite{s_second_1,s_second_2}. These questions are used for the model parameter estimation, where cancer status is the response variable. The third section includes four questions on the family's cancer status section, focusing on the participant's immediate family experience with the disease (\(q^3_1-q^3_4\)): did one of the participant's immediate family members have cancer and if so, what type and what were their relation to the participant, and how many immediate family members have been diagnostic with cancer. These questions are used to control for genetically related cancer properties \cite{s_third_1,s_third_2,s_third_3}. The fourth section includes eight questions (\(q^4_1-q^4_8\)) on the participant's stress and physiological signals of stress: daily-life stress, a belief that stress affects physical health, coping with stress, sleep difficulties, loss of appetite, fatigue, Trouble focusing, and stress impact on mental health. These questions are used to determine a single stress value for the participant \cite{s_last_1,s_last_2,s_last_3,s_last_4}. All questions in the stress section used a Likert scale \cite{joshi2015likert} from one to five, reflecting the least stress to the worst stress scale, respectively. 

Responses to the questionnaire were gathered using an online survey distributed across multiple channels, including email, social media, and community outreach programs. The questionnaire was designed to be inclusive, capturing responses from individuals across diverse socio-economic backgrounds. Participation was voluntary, with respondents informed about confidentiality and ethical considerations. Data collection was facilitated through Google Forms \cite{vasantha2016online}, which automatically anonymized responses to ensure privacy and adherence to ethical standards. The questionnaire remained open for four months, from December 2024 to March 2025.

\subsection{Analysis methods}
\label{sec:models}
We conduct three types of analysis that examine the dynamics in the data collected from the survey: statistical analysis, ML modeling, and causality modeling. Each of the analyses aims to capture a different property of the collected data. The statistical analysis aims to provide relatively simple yet straightforward descriptive statistics with correlations in the data and between-group analysis. The ML modeling ranges from computationally simple (decision tree) and explainable model to computationally complex (XGboost) and unexplainable model to study non-linear connections in the data. Finally, the causality modeling aims to capture the casual relationship between stress, immunity suppression, and cancer occurrence. For the analysis, We use Python (version 3.11) \cite{python}. The p-value for statistical significance was set at $p \leq 0.05$.

\subsubsection{Conceptual variables definitions}
We define the conceptual variables used in this study based on the survey questions. Formally, the cancer occurrence variable is binary and is based on the question "Have you ever been diagnosed with cancer?" in the survey. To obtain a single stress variable, we average the four stress questions in the survey's fourth section. This stress-level average is standardized to a value from 0 to 1 by subtracting one from the value and dividing it by five \cite{joshi2015likert}. The other four questions from the fourth section of the survey are used for the immunity suppression, following the same normalization process. All other questions in the survey are treated as individual variables in order to allow us to investigate the influence of each of these on the stress and cancer relationship, with or without immunity suppression.

\subsubsection{Statistical analysis}
\label{sec:fitting}
Three types of descriptive statistics are included in our analysis. Correlation analysis is performed using the Pearson correlation test \cite{pearson}. Between-group comparisons are performed using either a Student’s t-test or an ANalysis Of VAriance (ANOVA) with post-hoc t-tests and Bonferroni correction \cite{anova,t_test}. Regression analysis is performed by a logistic model utilizing the least mean squares method \cite{draper1998applied,lg_1}. The analysis results are presented as mean standard deviation (SD) for the continuous variables and as percentages for the categorical variables.

\subsubsection{Machine learning modeling}
We employed three machine learning models: Decision Trees (DT) \cite{dt_1}, XGBoost \cite{xgboost}, and k-Nearest Neighbors (KNN) \cite{knn}. Each model offers distinct benefits for the understanding and prediction of relationships in the dataset. DT offers interpretability andthe  ability to handle non-linearity while maintaining clear decision boundaries \cite{dt_2,dt_3}. XGBoost can capture complex interactions within the data, but is not explainable \cite{shmuel2024comprehensive}. KNN can predict an outcome - in our case, a cancer occurrence - for an individual, using the similarity of this individual to other individuals. To leverage this property, the KNN model is trained both on the entire variable set and on three sets of variables - only socio-demographic, and only stress combined with socio-demographic, and the model's performance was compared.

We divided the dataset into a training set  and a testing set that consisted of 80\% and 20\% of the data, respectively. This division ensures that the models have ample training data to learn from, while the separate test set allows for performance evaluation and assessment of general ization to new data \cite{carbonell1983overview}.
We employed 5-fold cross-validation to tune the models. This technique entails a split of the training data to five subsets. Each model is trained on four subsets and validated on the fifth. The process is repeated five times so that each subset serves as a validation set once \cite{wong2019reliable}. This technique mitigates issues related to variability in training data and ensures reliable performance metrics \cite{jung2018multiple}. Importantly, to ensure robust evaluation in clinical ML, the dataset was divided into training and validation cohorts with statistically similar age and sex distributions. This ensures the validation set reflects the training data. In our 5-fold cross-validation, all \(5\) subsets needed to maintain similar age and sex distributions \cite{ponv_6}.  The dataset was divided to five equal and distinct subsets such that the average distributional difference in age and sex between any two subsets is minimized. This optimization problem is conceptually related to the NP-hard nurse scheduling problem \cite{ponv_7,ponv_9}. We achieved a near-optimal solution using the Directed Bee Colony Optimization algorithm \cite{ponv_10,ponv_24}. A grid search \cite{alibrahim2021hyperparameter} for the optimal hyperparameters of each model was conducted. The grid search systematically evaluates combinations of hyperparameters to find the configuration that yields the best performance. The hyperparameters were the depth and minimum samples per split for the Decision Trees \cite{mantovani2018empirical}, the learning rate, maximum depth, and the number of trees for the XGBoost  \cite{glebov2023predicting}, and the number of neighbors and choice of distance metric (e.g., Euclidean or Manhattan distance) for the KNN \cite{rizki2024optimization}. 

The trained models are evaluated on the test set using standard classification metrics: accuracy, precision, recall, and \(F_1\) score \cite{Benchmarking_AutoML,Benchmarking_AutoML2}. To interpret the models and understand the contributions of different features, we employed the feature importance \cite{fi_1,fi_2,fi_3} and the Shapley Additive Explanations (SHAP) analysis \cite{Mayrhofer_Filzmoser_2023,meng2020makes}. The feature importance was implemented using the variable permutation method that evaluates the significance of each feature by randomly shuffling its values and measuring the corresponding drop in model performance. Features causing a greater reduction in performance are considered more important. The SHAP values provide a complementary interpretable measure of variable's impact by estimating each variable’s contribution to the model’s predictions. SHAP assigns positive or negative values to each feature, indicating whether they increase or decrease the likelihood of a particular outcome. 

\subsubsection{Causal modeling}
To establish causal relationships between stress, immunity suppression, and cancer occurrence, we implement the Rubin Causal Model (RCM) \cite{imbens2010rubin}. The RCM framework estimate treatment effects while accounting for potential confounders. The fundamental concept of RCM is the comparison between potential outcomes: the outcome if an individual receives a treatment versus the outcome if they do not. In our context, the \say{treatment} is defined as experiencing high-stress levels, and the outcome is cancer occurrence.

The formal definition of the model is as follows. let \( Y_i(1) \) and \( Y_i(0) \) represent the outcomes for individual \( i \) under treatment and control conditions, respectively. The treatment effect for individual \( i \) is defined as \(\tau_i := Y_i(1) - Y_i(0)\). Since we can only observe one of these outcomes per individual, we estimate the outcome by the average treatment effect (ATE) \( ATE := E[Y(1)] - E[Y(0)]\). To mitigate selection bias, we employ propensity score matching (PSM) \cite{benedetto2018statistical}. The propensity score \( e(X) \) represents the probability of an individual receiving treatment given their observed covariates \( X\) such that \(e(X) := P(T = 1 | X) \). We estimate \( e(X) \) using the logistic regression model ensuring that treatment and control groups are balanced in terms of covariates. Matching is then performed to compare treated and untreated individuals with similar propensity scores. After adjusting for confounders using the PSM, we estimated the treatment effect using regression models. Formally, we compare the adjusted means of the treatment group to the control group to derive causal conclusions. 

\section{Results}
\label{sec:results}

\subsection{Statistical analysis results}
This section includes three parts: descriptive statistics of the sample, followed by the results of the bivariate analysis, and then the multivariate analysis results.

A cohort of 1318 participants responded to the questionnaire, comprising 745 females (56.5\%), 566 males (43.0\%), and 7 individuals (0.5\%) who preferred not to disclose their gender. The participants' age range is 18 to 79 years, with a mean of 35.7 years and a standard deviation of 13.4. 
Most participants (820, 62.2\%) are married, 283 (21.5\%) are in a relationship, 214 (16.2\%) singles, 105 (8.0\%) divorced, and 48 (3.6\%) widow/ers. Most of the participants have two children (515, 39.1\%) followed by the groupsnof participants with no children (363, 27.5\%), one child (158, 12,0\%), three (101, 7.6\%), four (93, 7.0\%), and five or more (88, 6.6\%) children. 
Regarding education, 89 (6.8\%) participants did not complete high school, 267 (20.3\%) completed high school, 426 (32.3\%) hold a Bachelor's degree, 368 (27.9\%) hold a Master's degree, and 168 (12.7\%) hold a PhD or M.D. Participants reported living with a median of three other individuals in their household, where 231 (17.5\%) live alone, 329 (25.0\%) live with one other person, 416 (31.5\%) live with two others, 227 (17.2\%) live with three others, and 115 (8.7\%) live with four or more cohabitants. In terms of household income, 198 participants (15.0\%) reported earning up to 2,150 USD per month. A total of 134 participants (10.2\%) reported incomes between 2,150–2,700 USD, 143 (10.8\%) between 2,700–3,250 USD, 239 (18.1\%) between 3,250–5,400 USD, 219 (16.6\%) between 5,400–8,000 USD, and 137 (10.4\%) between 8,100–10,800 USD. An additional 101 participants (7.7\%) reported earning over above 10,800 USD, and 147 (11.1\%) chose not to disclose their income. Over 89\%  of the participants reside in one of the following five countriesthe United States (281 participants; 21.3\%), Israel (255; 19.3\%), Italy (247; 18.7\%), the United Kingdom (226; 17.1\%), and Russia (166; 12.6\%). The remaining 143 participants (10.8\%) were distributed across other 29 countries. 
Of the participants, 237 (18.0\%) were or are currently diagnosed with cancer. 

\begin{figure}[!ht]
    \centering
    \includegraphics[width=0.99\linewidth]{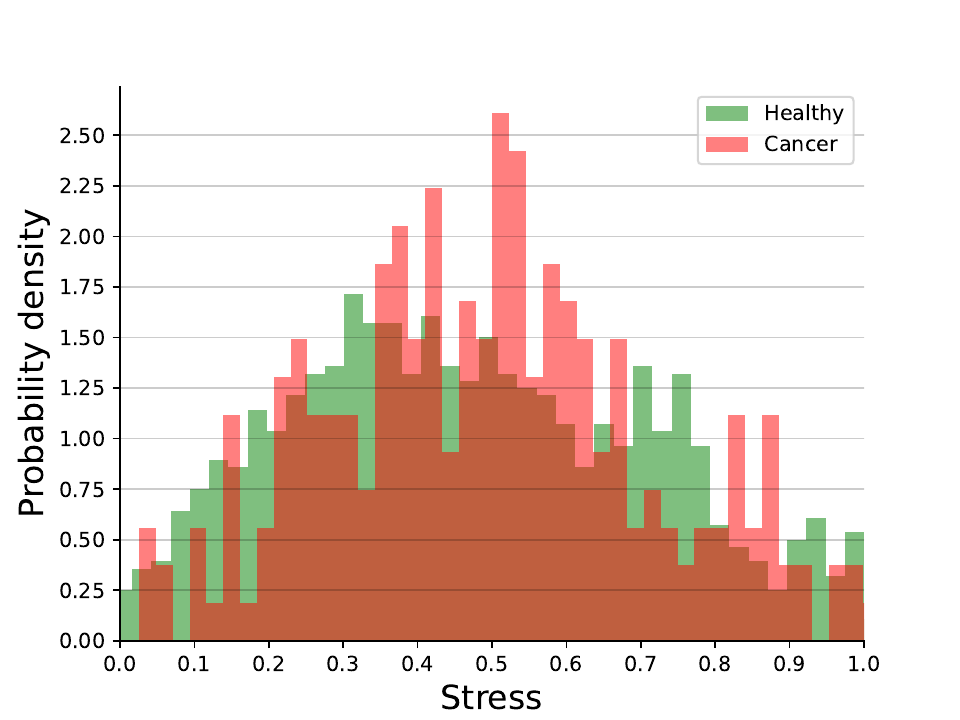}
    \caption{Histogram of the stress level of the participants divided into those who had cancer and those who did not.}
    \label{fig:hist}
\end{figure}

Fig. \ref{fig:hist} shows the histogram of the participants' stress levels, divided into participants who had cancer and those who did not. Initially, using the Shapiro-Wilk test \cite{hanusz2014simulation}, we tested if the data is normally distributed, obtaining that stress levels for both healthy and cancer individual are not normally distributed with \(p = 0.008\) and \(p = 0.023\), respectively. Using a Mann–Whitney U test, we show there is no statistical difference between the two distributions with \(p = 0.108\). 

Next, we computed the univariate statistical relationship between the variables with stress levels and the occurrence of cancer. For the stress level, we assume it is continuous (due to the 625 possible values it may get and the sample size) and used Person correlation while for the cancer occurrence, we used the \(\chi^2\) test. Starting with the stress levels, cancer treatment received a relatively high association with stress with 0.32 \((p < 0.01)\). Moreover, individuals with a higher number of cancer cases in the family are usually more stressed (0.21, \(p < 0.05\)).  Furthermore, stress frequency was found to be moderately correlated to stress levels (0.53, \(p < 0.01\)), followed by the belief that stress impacts one's health (0.33, \(p < 0.05\)). In a similar manner, the self-evaluation of one's ability to cope with stress has a relatively minor association with self-reported stress levels (-0.13, \(p < 0.05\)). Among the demographic variables, participants' age and number of children show a small positive correlation with stress with 0.07 \((p < 0.01)\) and 0.09 \((p < 0.05)\), respectively, as well as education level (0.04, \(p < 0.05\)). The other parameter do not show a statistically significant correlation with stress. Next, the analysis revealed a robust association between the history of cancer in the family and cancer occurrence \(\chi^2 = 62.3, p < 0.01\), while no other parameters show statistically significant relationships. 

Table \ref{table:lr} presents the results of a logistic regression analysis examining factors associated with the occurrence of cancer. Several variables demonstrated statistically significant associations with the odds of a cancer occurrence. Specifically, older age \(\beta = 0.090, p = 0.002\), a family history of cancer \(\beta = 0.128, p = 0.003\), higher number of family members with cancer \(\beta = 0.104, p = 0.019\), higher stress frequency \(\beta = 0.031, p = 0.045\), greater perceived health impact of stress \(\beta = 0.062, p = 0.047\), and higher general stress levels \(\beta = 0.091, p = 0.034\) were all positively associated with increased odds of cancer occurrence. While the estimate for higher education level was positive \(\beta = 0.017\), it did not reach statistical significance (\(p = 0.094\)). Similarly, gender, marital status, number of children, income, household size, stress-coping ability, sleep issues, appetite loss, and fatigue did not show statistically significant associations with cancer occurrence in this model.

\begin{table}[h]
\centering
\begin{tabular}{clccc}
\hline \hline
\textbf{Q-Index} & \textbf{Parameter} & \textbf{Estimate} & \textbf{p-value} & \textbf{95\% CI} \\
 \hline \hline
1 & Gender (Male) & -0.038 & 0.409 & [-0.128, 0.052] \\
2 & Age & 0.090 & 0.002 & [ 0.031, 0.149] \\
3 & Marital Status (Ref: Single) & -0.080 & 0.151 & [-0.189, 0.029] \\
4 & Children & -0.028 & 0.277 & [-0.077, 0.021] \\
5 & Education Level (Ref: Low) & 0.017 & 0.094 & [-0.003, 0.037] \\
6 & Income & 0.069 & 0.149 & [-0.025, 0.163] \\
7 & Household Size & 0.060 & 0.315 & [-0.058, 0.178] \\
8 & Family Cancer History (Yes) & 0.128 & 0.003 & [ 0.042, 0.214] \\
9 & Family Cancer Count & 0.104 & 0.019 & [ 0.017, 0.191] \\
10 & Stress Frequency & 0.031 & 0.045 & [ 0.001, 0.061] \\
11 & Stress Health Impact & 0.062 & 0.047 & [ 0.001, 0.123] \\
12 & Stress Coping Ability & -0.033 & 0.067 & [-0.069, 0.003] \\
13 & Sleep Issues & 0.021 & 0.483 & [-0.040, 0.082] \\
14 & Appetite Loss & -0.018 & 0.253 & [-0.050, 0.014] \\
15 & Fatigue & 0.012 & 0.070 & [-0.001, 0.025] \\
16 & Concentration Issues & 0.032 & 0.186 & [-0.048, 0.112] \\
17 & Stress (General) & 0.091 & 0.034 & [ 0.007, 0.175] \\ \hline \hline
\end{tabular}
\caption{Logistic regression results for cancer occurrence. Notably, marital status and level of education are treated as categorical variables with the reference category indicated in parentheses.}
\label{table:lr}
\end{table}

\subsection{Machine learning results}
Table \ref{tab:model_performance} presents the performance metrics of the ML models used for predicting cancer occurrence based on socio-demographic, stress-related, and immune suppression factors. The XGBoost-based model achieved the highest performance with an accuracy of 0.73 and an F1 score of 0.68. These values imply close to optimal predictive capabilities on the proposed data, as XGboost is considered the SOTA on tabular data \cite{shmuel2024comprehensive}. The DT-based model performed moderately well, but significantly worse compared to the XGboost, with an accuracy of 0.64 and an F1-score of 0.60. The KNN classifier was tested on different feature sets. When using all variables, KNN achieved an accuracy of 0.68 and an F1 score of 0.66, showing better performance compared to the DT model while worse compared to the XGboost model. When trained on only socio-demographic variables, stress-related variables, or immune suppression factors, its performance marginally declined. The latter two were just slightly better than a random guess (which will produce a 0.5 score). 

The XGBoost model's superior accuracy highlight its ability to effectively capture complex interactions within the data and suggests that the interactions among socio-demographic, stress-related, and immune suppression variables are intricate and nonlinear. The Decision Tree (DT) model, a simpler- hierarchical splits, and global model that offers the advantage of being interpretable, demonstrated lower performance compared to XGBoost. This performance may indicate that the relationships in the data are not  easily captured through straightforward hierarchical splits. The k-Nearest Neighbors (KNN) classifier, which showed moderate performance when using all variables is an interesting contrast. Its performance superiority over DT yet inferiority to XGBoost implies that while KNN takes advantage of the proximity-based relationships in the data, it may struggle with scalability and dimensionality compared to XGBoost. The decline in KNN's performance when restricted to subsets of variables (socio-demographic, stress-related, or immune suppression) may indicate that cancer occurrence is likely influenced by a combination of factors rather than isolated variable groups. The relatively better performance for the socio-demographic variables subset may indicate, howvere, that cancer occurrence dynamics are directly associated with socio-economic properties but are not directly with stress levels and immune suppression. 

\begin{table}[h]
    \centering
    \begin{tabular}{lcccc}
        \hline \hline
        \textbf{Model} & \textbf{Accuracy} & \textbf{F1 Score} & \textbf{Recall} & \textbf{Precision} \\
        \hline \hline
        XGboost & 0.73 & 0.68 & 0.64 & 0.72 \\
        DT & 0.64 & 0.60 & 0.57 & 0.63 \\
        KNN (all variables) & 0.68 & 0.66 & 0.63 & 0.69 \\
        KNN (only socio-demographic) & 0.61 & 0.59 & 0.53 & 0.66 \\
        KNN (only stress) & 0.54 & 0.53 & 0.55 & 0.51 \\
        KNN (only immune suppression) & 0.56 & 0.55 & 0.56 & 0.53 \\
        \hline \hline
    \end{tabular}
    \caption{Performance metrics for different models}
    \label{tab:model_performance}
\end{table}

\begin{figure}[!ht]
    \centering
    
    \begin{subfigure}{.49\textwidth}
    \includegraphics[width=0.99\textwidth]{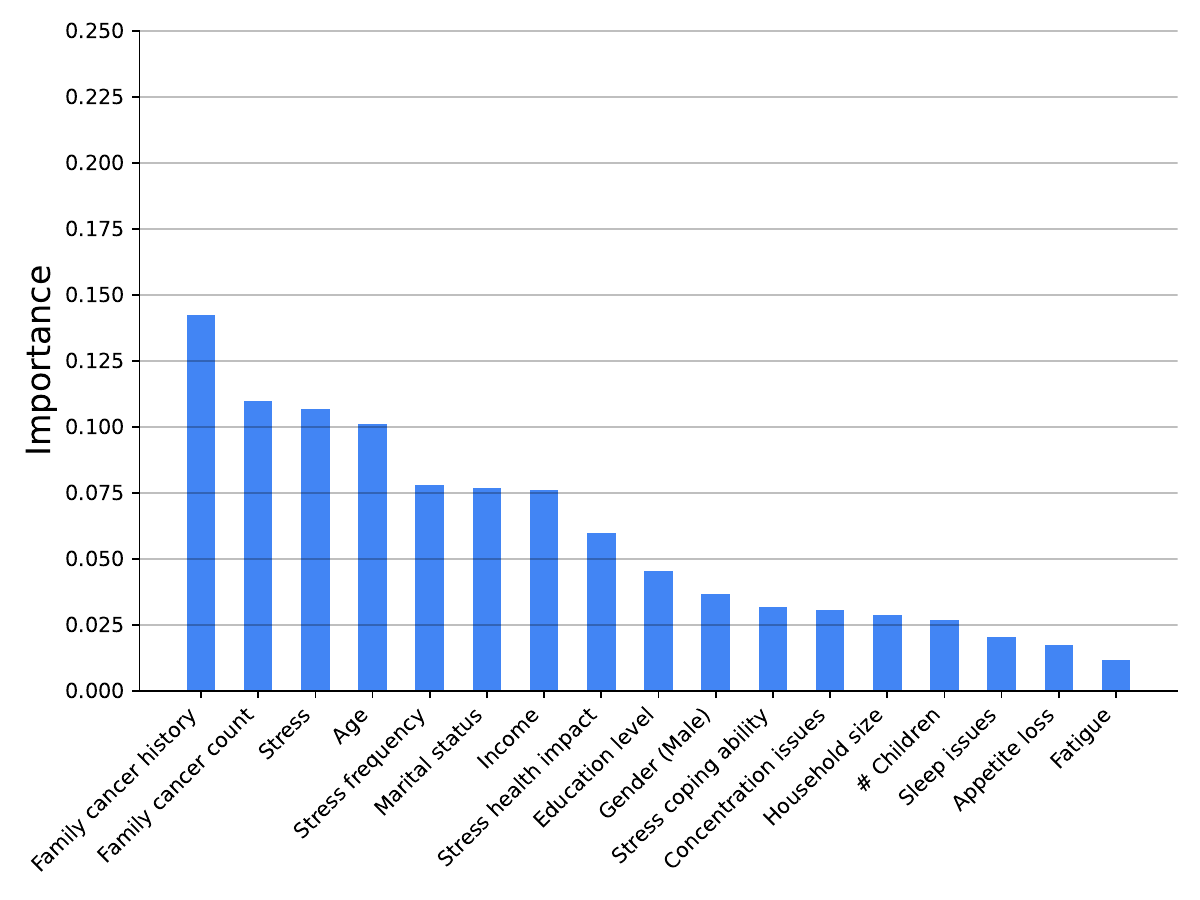}
        \caption{XGboost.}
        \label{fig:fi_xgboost}
    \end{subfigure}
    \begin{subfigure}{.49\textwidth}
    \includegraphics[width=0.99\textwidth]{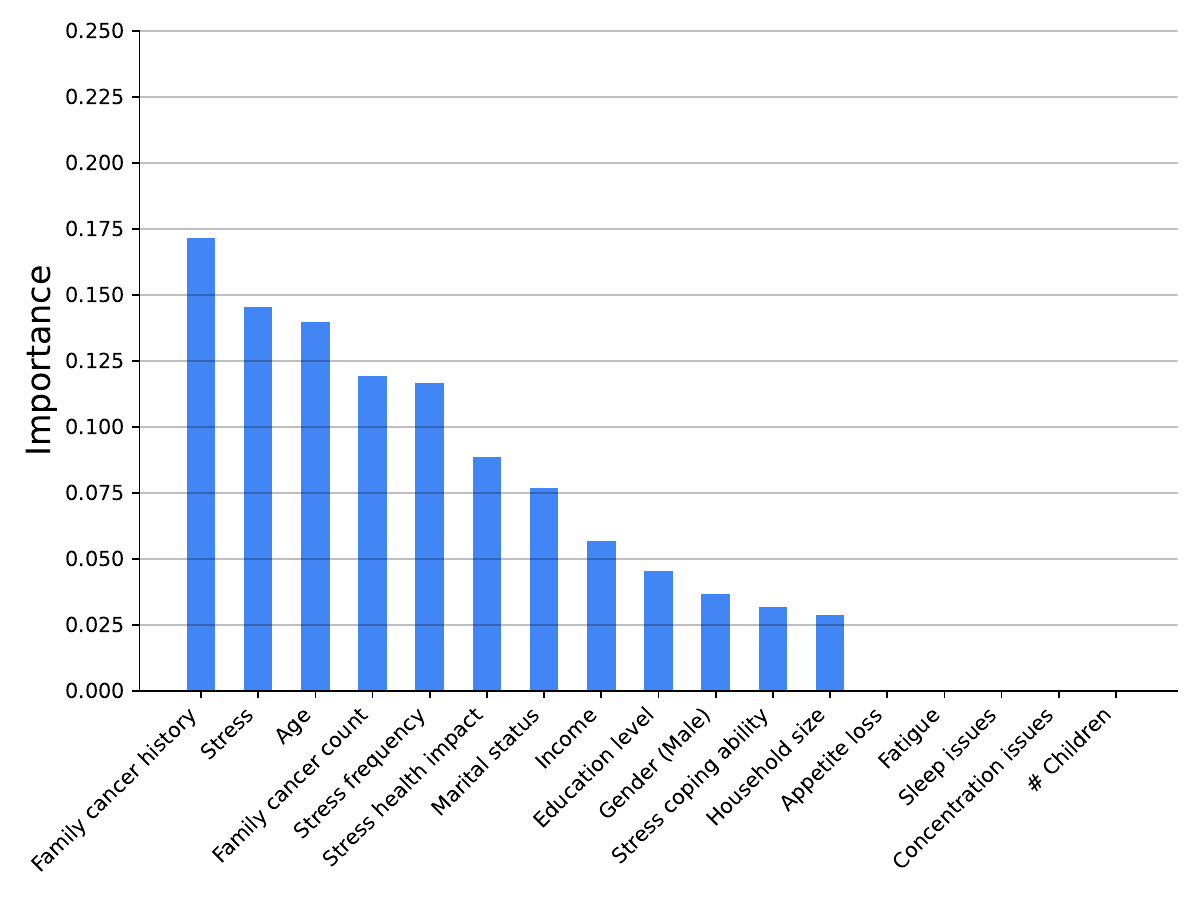}
        \caption{Decision Tree.}
        \label{fig:fi_dt}
    \end{subfigure}
    
    \begin{subfigure}{.49\textwidth}
    \includegraphics[width=0.99\textwidth]{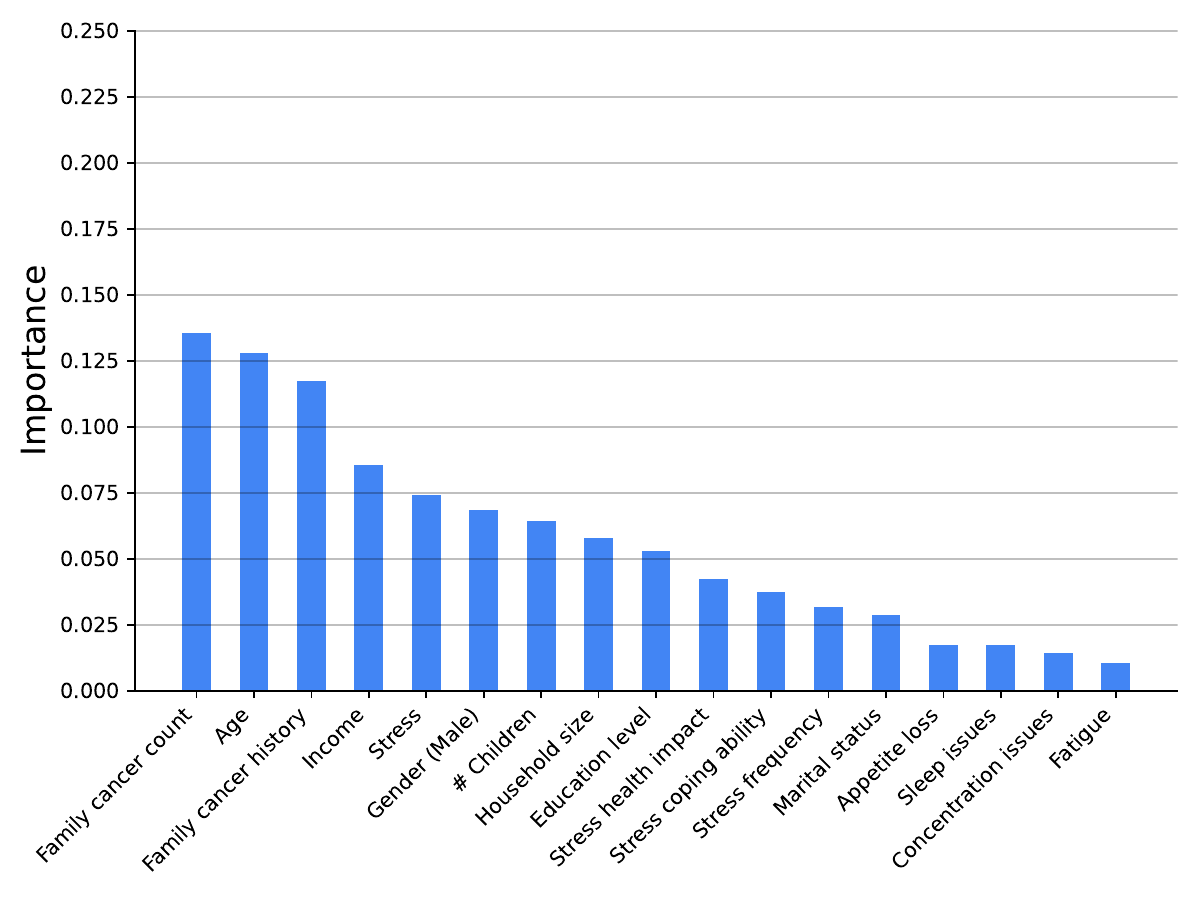}
        \caption{KNN (all).}
        \label{fig:fi_knn_all}
    \end{subfigure}
    \begin{subfigure}{.49\textwidth}
    \includegraphics[width=0.99\textwidth]{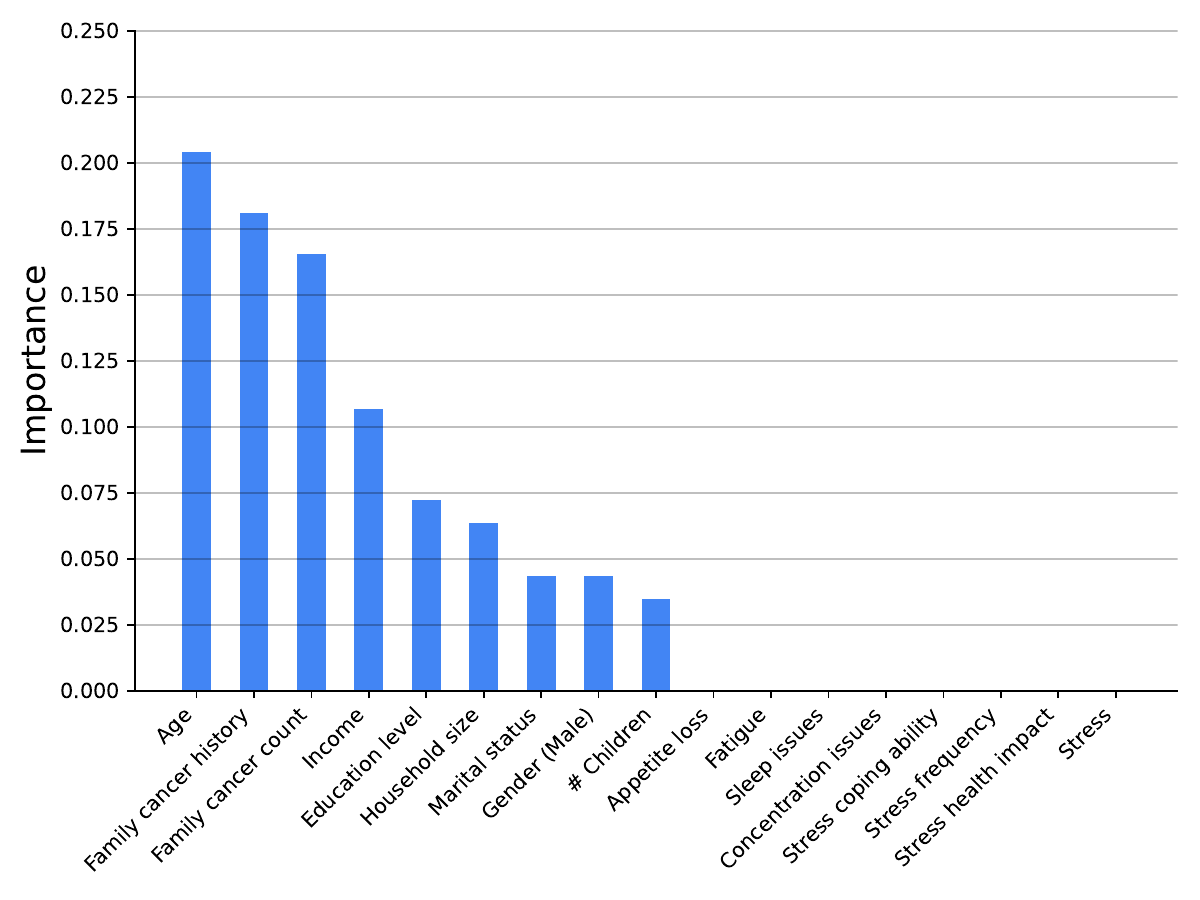}
        \caption{KNN (only socio-demographic).}
        \label{fig:fi_knn_socio}
    \end{subfigure}
    
    \begin{subfigure}{.49\textwidth}
    \includegraphics[width=0.99\textwidth]{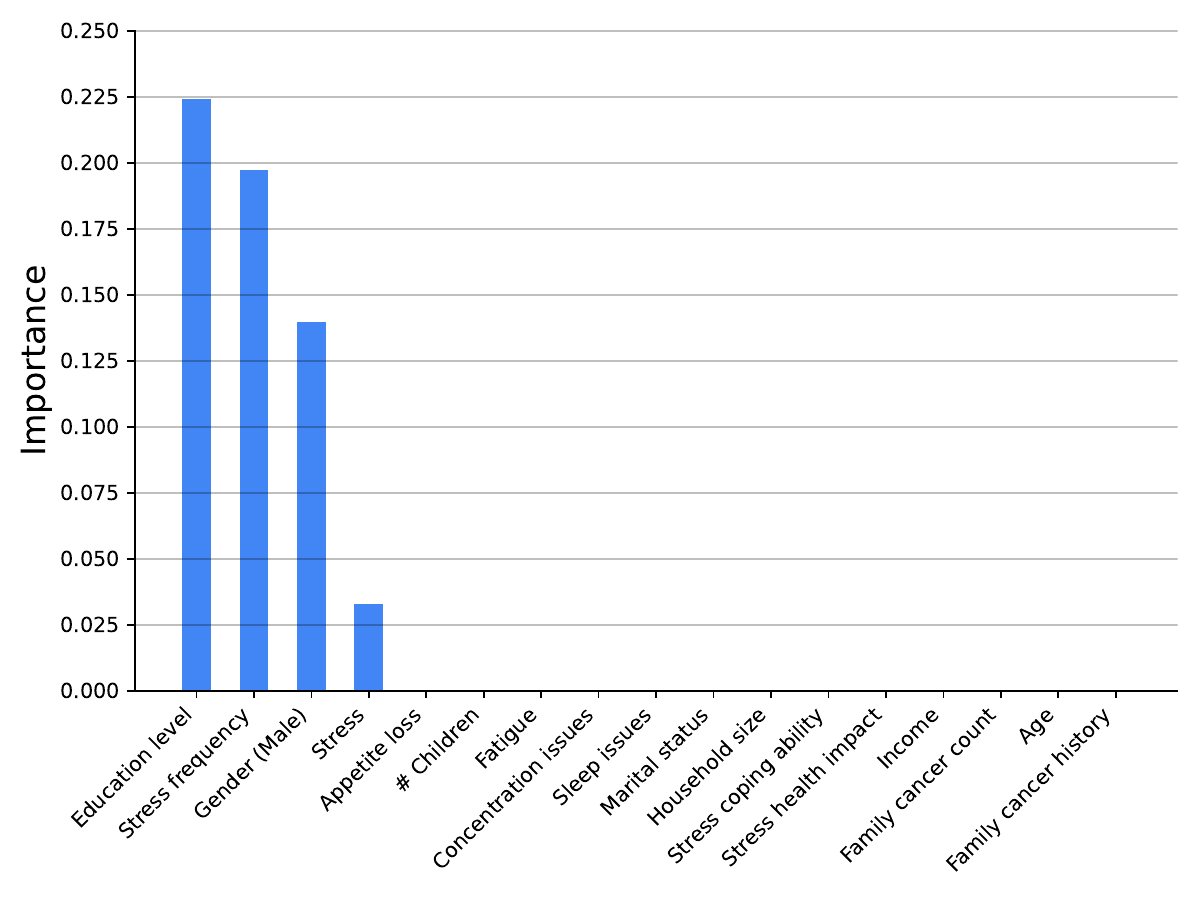}
        \caption{KNN (only stress).}
        \label{fig:fi_knn_stress}
    \end{subfigure}
    \begin{subfigure}{.49\textwidth}
    \includegraphics[width=0.99\textwidth]{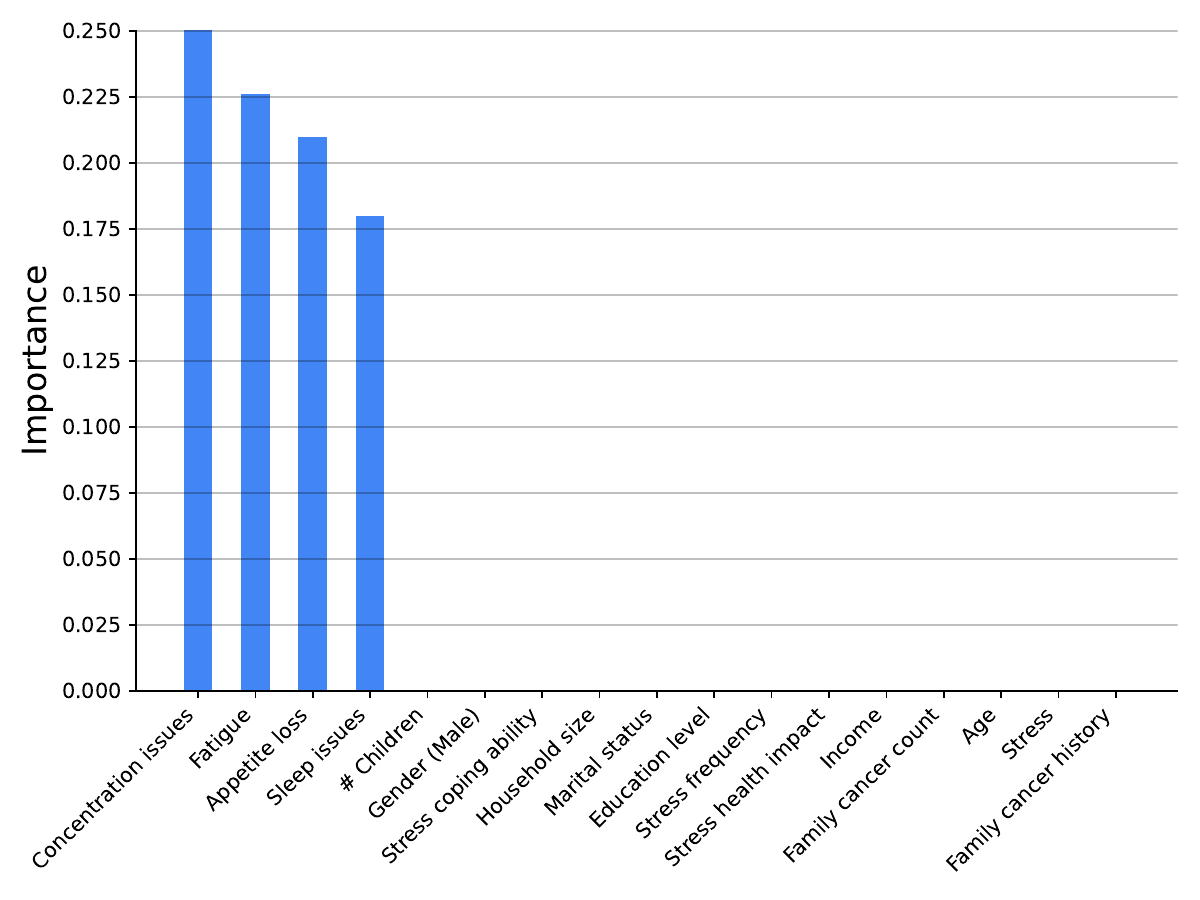}
        \caption{KNN (only immune suppression).}
        \label{fig:fi_knn_immune}
    \end{subfigure}
    \caption{The feature importance distribution of the six machine learning models predicting cancer occurrence. }
\label{fig:ml_fi}
\end{figure}

Fig. \ref{fig:ml_fi} shows the variable's importance analysis for the six obtained ML models. For all the models that include family cancer history, it is the most important variable. Age is also found to be important across the models, being the second most important variable in KNN model as well as fourth and third for the XGboost and DT models, respectively. For the (global) tree-based models, the XGboost and DT, the level of stress is the third and second most important variable, respectively, while for the (local) similarity-based KNN model, stress is already the fifth most important variable. 

\subsection{Casual model results}
For the casual model, we adopt a two-phase analysis: first, we analyze a direct causal effect between all the model's variables and cancer occurrence; second, we assume a specific causal graph based on the literature. This approach allows us to obtain a simple yet straightforward causal relationship on one hand and a more complex and theory-driven one on the other hand.  

\begin{figure}[!ht]
    \centering
    \begin{tikzpicture}[->,>=stealth, node distance=0.3cm]
        \node (gender) [rectangle, draw] {Gender};
        \node (age) [rectangle, draw, below=of gender] {Age};
        \node (marital) [rectangle, draw, below=of age] {Marital Status};
        \node (children) [rectangle, draw, below=of marital] {Children};
        \node (edu) [rectangle, draw, below=of children] {Education Level};
        \node (income) [rectangle, draw, below=of edu] {Income};
        \node (household) [rectangle, draw, below=of income] {Household Size};
        \node (famHist) [rectangle, draw, below=of household] {Family Cancer History};
        \node (famCount) [rectangle, draw, below=of famHist] {Family Cancer Count};
        \node (stressFreq) [rectangle, draw, below=of famCount] {Stress Frequency};
        \node (stressImpact) [rectangle, draw, below=of stressFreq] {Stress Health Impact};
        \node (stressCoping) [rectangle, draw, below=of stressImpact] {Stress Coping Ability};
        \node (sleep) [rectangle, draw, below=of stressCoping] {Sleep Issues};
        \node (appetite) [rectangle, draw, below=of sleep] {Appetite Loss};
        \node (fatigue) [rectangle, draw, below=of appetite] {Fatigue};
        \node (concentration) [rectangle, draw, below=of fatigue] {Concentration Issues};
        \node (stressGeneral) [rectangle, draw, below=of concentration] {General Stress};
        \node (cancer) [rectangle, draw, right=8cm of famHist] {Cancer Occurrence};
        
        \draw (gender) -- node[above] {0.018} (cancer);
        \draw (age) -- node[above] {0.103**} (cancer);
        \draw (marital) -- node[above] {0.030} (cancer);
        \draw (children) -- node[above] {0.045} (cancer);
        \draw (edu) -- node[above] {0.047} (cancer);
        \draw (income) -- node[above] {0.079} (cancer);
        \draw (household) -- node[above] {0.060} (cancer);
        \draw (famHist) -- node[above] {0.138**} (cancer);
        \draw (famCount) -- node[above] {0.104*} (cancer);
        \draw (stressFreq) -- node[above] {0.031*} (cancer);
        \draw (stressImpact) -- node[above] {0.062*} (cancer);
        \draw (stressCoping) -- node[above] {0.033} (cancer);
        \draw (sleep) -- node[above] {0.021} (cancer);
        \draw (appetite) -- node[above] {0.018} (cancer);
        \draw (fatigue) -- node[above] {0.012} (cancer);
        \draw (concentration) -- node[above] {0.032} (cancer);
        \draw (stressGeneral) -- node[above] {0.091*} (cancer);
    \end{tikzpicture}
    \caption{Path diagram of direct casual model between the features and cancer occurrence. Standardized path coefficients and significance levels are indicated (\(^*\; p < 0.05, \;\; ^{**} \; p < 0.01\)).}
    \label{fig:causal_model}
\end{figure}
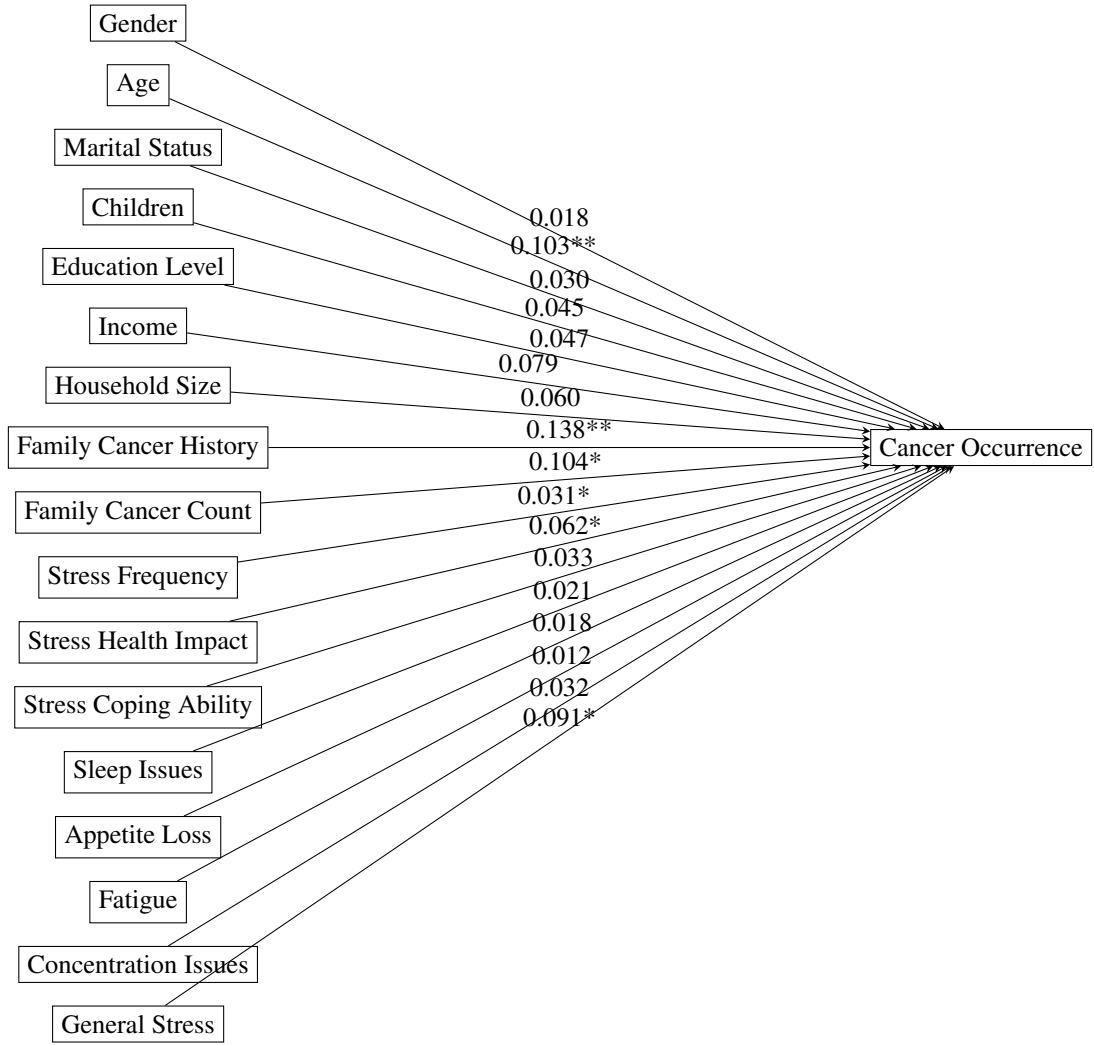

Figure \ref{fig:causal_model} presents the causal model for predicting cancer occurrence, assuming a direct effect of all variables. Notably, age (0.103, \(p < 0.01\)) and family cancer history (0.138, \(p < 0.01\)) exhibit the strongest associations with cancer occurrence, highlighting their critical role in risk assessment. Other significant contributors include family cancer count (0.104, \(p < 0.05\)), stress frequency (0.031, \(p < 0.05\)), stress health impact (0.062, \(p < 0.05\)), and stress levels (0.091, \(p < 0.05\)), underscoring the interplay between genetic predisposition and psychological stress in cancer risk. These results suggest that while hereditary factors remain dominant, psychosocial stressors may also play a meaningful role in cancer susceptibility.

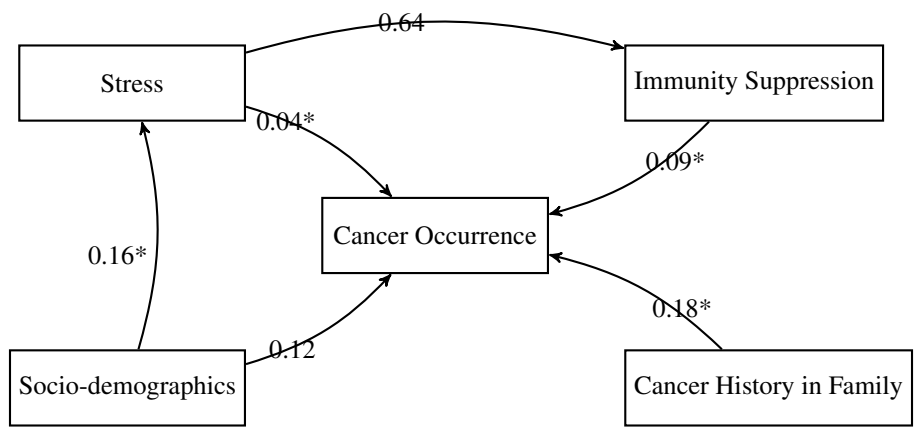
\begin{figure}[h]
    \centering
    \begin{tikzpicture}[->,>=stealth', node distance=4cm, thick,
                        box/.style={draw, rectangle, minimum width=3cm, minimum height=1cm}]  
        \node (C) [box] {Cancer Occurrence};

        \node (S) [box, above left=1cm and 1cm of C] {Stress};  
        \node (I) [box, above right=1cm and 1cm of C] {Immunity Suppression}; 
        \node (FH) [box, below right=1cm and 1cm of C] {Cancer History in Family}; 
        \node (SD) [box, below left=1cm and 1cm of C] {Socio-demographics}; 

        \draw [->] (S) to [bend left=15] node[above left] {0.04*} (C);
        \draw [->] (I) to [bend left=15] node[above right] {0.09*} (C);
        \draw [->] (FH) to [bend right=15] node[below right] {0.18*} (C);
        \draw [->] (SD) to [bend right=15] node[below left] {0.12} (C);

        \draw [->] (S) to [bend left=15] node[left] {0.64} (I);

        \draw [->] (SD) to [bend right=15] node[below left] {0.16*} (S); 
        
    \end{tikzpicture}
    \caption{Path diagram of hypothesized relationships influencing cancer occurrence. Standardized path coefficients and significance levels are indicated (\(^*\; p < 0.05\)).}
    \label{fig:casual_2}
\end{figure}

Figure \ref{fig:casual_2} presents a hypothesized model for the relationships between stress and cancer occurrence while controlling for potential confounding factors - socio-demographics, immune suppression, and cancer history in the family. Specifically, the model posits that stress may directly influence cancer occurrence ($\beta = 0.04, p < 0.05$), as well as indirectly through immunity suppression ($\beta = 0.64$), which in turn, impacts cancer occurrence ($\beta = 0.09, p < 0.05$). The model further incorporates cancer history in family ($\beta = 0.18, p < 0.05$) and socio-demographics as direct predictors of cancer occurrence, allowing us to assess their independent contributions. Finally, socio-demographics are also hypothesized to influence stress levels ($\beta = 0.16, p < 0.05$), acknowledging the potential role of social factors in shaping stress responses. 

\section{Discussion and Conclusion}
\label{sec:discussion}
Our study explores the contribution of chronic stress and immune suppression to cancer occurrence. Insights from this association have significant implications for precision medicine strategies aimed at mitigating stress-induced changes in immune function \cite{intro_5_1,Antoni2006TheIO}. Quantifying and understanding this association may help in developing interventions to reduce cancer risk and improve treatment outcomes \cite{intro_11}. 

Our results offer a nuanced understanding of cancer risk factors. They integrate survey data from four key categories of variables, which were hypothesized as related to cancer occurrence: stress, socio-demographic factors, family cancer history, and immune suppression. 

Our machine learning models effectively capture the interactions between variables. These findings reinforce the significance of stress in cancer prediction, and highight the potential of machine learning in capturing and modeling cancer risk \cite{intro_15_1,intro_15_2,glebov2023predicting,intro_17_2}. Specifically, the combination and comparison of different ML models is demonstrated as effective, where the better prediction accuracy of XGBoost was combined with the explainability of KNN.  

Our traditional statistical methods corroborate these insights and indicate significant correlations between stress variables and cancer risk \cite{Cohen2001PsychologicalSA,Butow2018DoesSI}. 

However, our causal model extends beyond correlation and provides insights into how these factors interact and contribute to cancer risk. 
The most important findings of the study were the identified causal relationships between both variables categories (figure 5) and individual variables (figure 4) and cancer occurrence. Age and family cancer history (figure 4) are revealed as dominant factors in cancer risk, corroborating the importance of hereditary influences \cite{intro_2,s_third_1}.
Psychosocial stressors, including stress frequency, health impact, and stress levels, also play a significant role, as suggested previously \cite{Mohan2022PsychosocialSA,Antoni2006TheIO}. 
Our model (figure 5) suggests that both stress and immunity suppression directly influences cancer. The indirect effect of stress on cancer occurrence through immune suppression was not significant according to our causal model \cite{segerstrom2004psychological,Dhabhar2013PsychologicalSA}. These findings add depth to existing research by quantifying these relationships and their interconnections.
The significant correlations between stress frequency, stress health impact, and cancer occurrence suggest that individuals who perceive stress as significantly affecting their health and experience frequent stress might be at increased risk for cancer \cite{Dai2020ChronicSP,Mohan2022PsychosocialSA}. This could be due to stress-related behavioral changes and physiological effects impacting cancer risk \cite{segerstrom2004psychological}.
The significant correlation with age was previously associated with immunosenescence, which exhibits similarities to stress-induced immune alterations. Elderly individuals frequently display elevated cortisol levels and hypothalamic-pituitary-adrenal axis activation, potentially exacerbating stress-related immune dysfunction \cite{Cohen2001PsychologicalSA,Bauer2009TheRO}. Older age may cause cancer risk due to cumulative exposure to environmental risk factors over time and the natural accumulation of genetic mutations \cite{intro_2,intro_3}.
These insights are valuable additions to the existing body of epidemiological studies that displays contradicting evidence of stress correlation to cancer risk \cite{Heikkil2013WorkSA,Butow2018DoesSI,Lempesis2023RoleOS}. 

The findings suggest actionable insights to cancer risk mitigation. The causal effects of stress variables reflect chronic stress exposure, the way individuals perceive and experience stress, unhealthy coping mechanisms such as poor diet, lack of exercise, and substance use \cite{MorenoSmith2010ImpactOS}. This suggests that interventions directly targeting stress management might be effective as cancer prevention strategies.

Our methodology entails several limitations. The reliance on self-reported data introduces potential biases, affecting the precision of stress and health assessments \cite{CortsIbez2022TheVO}. Online self-recruited participants may limit the generalizability of the findings to populations that don't have internet access or are underrepresented - individuals with low socioeconomic status or those in rural areas - in online surveys \cite{qu_2}. Specific questions - history of cancer diagnosis, stress-related experiences - are vulnerable to recall bias, where participants might inaccurately recall past events or experiences \cite{Butow2018DoesSI}. 
The survey variables of stress and immune suppression may oversimplify the nuanced physiological and psychological dimensions of chronic stress and immune function, compared to objective biomarkers such as stress-related hormones like cortisol or immune-related cytokines \cite{Antoni2006TheIO,Dhabhar2013PsychologicalSA,Nakata2012PsychosocialJS}. The study cross-sectional design limits the ability to draw definitive causal conclusions \cite{Butow2018DoesSI}, and longitudinal data is needed \cite{Pham2024ChronicSR}. Moreover, while causal modeling can control for observed confounders, unmeasured variables - comorbidities, specific lifestyle factors like smoking or alcohol consumption - may still confound the observed relationships. This is particularly relevant to stress and cancer risk as both may be influenced by hidden shared factors \cite{Butow2018DoesSI,Avgerinos2019ObesityAC}.

Taken jointly, our study quantifies the multidimensional nature of cancer risk \cite{MorenoSmith2010ImpactOS}, and the combined effect of genetic predisposition, stress perception, and socio-demographic factors. 
Longitudinal studies should explore these relationships over time to better understand their dynamics and potential for intervention. Our study shows the way in this context, as it reinforces the need for approaches that integrate diverse data types to capture the full spectrum of influences on cancer occurrence \cite{intro_14}. Our findings highlight the benefit of integrating machine learning and causal models in the discovery of patterns that traditional and simpler models might miss \cite{intro_15_1}. These patterns provide greater insight into complex health datasets of cancer, and enhance the ability to develop predictive models for cancer occurrence \cite{intro_15_2,intro_14}. By identifying modifiable stress-related factors and quantifying their impact on cancer occurrence, these models can inform targeted prevention strategies and enhance early screening protocols tailored to high-risk individuals \cite{Mohan2022PsychosocialSA,Pham2024ChronicSR,Bahri2019TheRB}. 
Addressing psychological stress through interventions such as stress reduction techniques, lifestyle modification programs, and psycho-oncological support has been shown to effectively reduce cancer risk when integrated into broader prevention frameworks \cite{Dai2020ChronicSP,Antoni2006TheIO}. Evidence from predictive models could guide the development of public health interventions aimed at reducing the global cancer burden.

\section*{Declarations}


\subsection*{Consent for publication}
Not Applicable

\subsection*{Availability of data and materials}
The study's data is available upon reasonable request from the corresponding author.

\subsection*{Competing Interests}
The authors declare no competing interests.

\subsection*{Funding}
This study received no funding.

\subsection*{Acknowledgements}
The authors wish to thank Eden Aloni for suggesting the research question at the core of this study. 

\subsection*{Authors' contributions}
Teddy Lazebnik: Conceptualization, Methodology, Software, Formal analysis, Investigation, Data Curation, Writing - Original Draft, Writing - Review \& Editing, Visualization. \\
Vered Aharonson: Validation, Investigation, Writing - Original Draft, Writing - Original Draft, Writing - Review \& Editing.

\bibliography{biblio}

\begin{thebibliography}{100}

\bibitem{intro_1}
F.~Bray, M.~Laversanne, E.~Weiderpass, and I.~Soerjomataram.
\newblock The ever-increasing importance of cancer as a leading cause of premature death worldwide.
\newblock {\em Cancer}, 127(16):3029--3030, 2021.

\bibitem{intro_2}
B.~A.~J. Ponder.
\newblock Cancer genetics.
\newblock {\em Nature}, 411(6835):336--341, 2001.

\bibitem{intro_3}
P.~Boffetta and F.~Nyberg.
\newblock Contribution of environmental factors to cancer risk.
\newblock {\em British medical bulletin}, 68(1):71--94, 2003.

\bibitem{intro_4}
S.~K. Lutgendorf and A.~K. Sood.
\newblock Biobehavioral factors and cancer progression: physiological pathways and mechanisms.
\newblock {\em Psychosomatic medicine}, 73(9):724--730, 2011.

\bibitem{intro_5_1}
J.~Z. Borysenko.
\newblock Behavioral-physiological factors in the development and management of cancer.
\newblock {\em General Hospital Psychiatry}, 4(1):69--74, 1982.

\bibitem{intro_11}
M.~Chen, A.~K. Singh, and E.~A. Repasky.
\newblock Highlighting the potential for chronic stress to minimize therapeutic responses to radiotherapy through increased immunosuppression and radiation resistance.
\newblock {\em Cancers}, 12(12):3853, 2020.

\bibitem{Mohan2022PsychosocialSA}
Ananyaa Mohan, Inge Huybrechts, and Nathalie Michels.
\newblock Psychosocial stress and cancer risk: a narrative review.
\newblock {\em European Journal of Cancer Prevention}, 31:585 -- 599, 2022.

\bibitem{intro_6}
R.~Glaser and J.~Kiecolt-Glaser.
\newblock Stress damages immune system and health.
\newblock {\em Discovery medicine}, 5(26):165--169, 2009.

\bibitem{intro_7}
A.~d’Onofrio, F.~Gatti, P.~Cerrai, and L.~Freschi.
\newblock Delay-induced oscillatory dynamics of tumour--immune system interaction.
\newblock {\em Mathematical and Computer Modelling}, 51(5-6):572--591, 2010.

\bibitem{Butow2018DoesSI}
Phyllis~N. Butow, Melanie~Anne Price, Joseph~R Coll, Katherine~L. Tucker, Bettina Meiser, Roger~L. Milne, Judy Wilson, Louise Heiniger, Brandi Baylock, Tracey Bullen, Prue~C Weideman, and Kelly-Anne Phillips.
\newblock Does stress increase risk of breast cancer? a 15‐year prospective study.
\newblock {\em Psycho‐Oncology}, 27:1908 -- 1914, 2018.

\bibitem{intro_14}
M.~Shehab, L.~Abualigah, Q.~Shambour, M.~A. Abu-Hashem, M.~K.~Y/ Shambour, A.~I. Alsalibi, and A.~H. Gandomi.
\newblock Machine learning in medical applications: A review of state-of-the-art methods.
\newblock {\em Computers in Biology and Medicine}, 145:105458, 2022.

\bibitem{intro_15_1}
M.~Levi, T.~Lazebnik, S.~Kushnir, N.~Yosef, and D.~Shlomi.
\newblock Machine learning computational model to predict lung cancer using electronic medical records.
\newblock {\em Cancer Epidemiology}, 92:102631, 2024.

\bibitem{intro_15_2}
M.~Amine Naji, S.~El~Filali, K.~Aarika, EL~H. Benlahmar, R.~A. Abdelouhahid, and O.~Debauche.
\newblock Machine learning algorithms for breast cancer prediction and diagnosis.
\newblock {\em Procedia Computer Science}, 191:487--492, 2021.

\bibitem{intro_16_1}
D.~Benrimoh, A.~Kleinerman, T.~A. Furukawa, C.~F. Reynolds~III, E.~J. Lenze, J.~Karp, B.~Mulsant, C.~Armstrong, J.~Mehltretter, R.~Fratila, et~al.
\newblock Towards outcome-driven patient subgroups: a machine learning analysis across six depression treatment studies.
\newblock {\em The American Journal of Geriatric Psychiatry}, 32(3):280--292, 2024.

\bibitem{intro_16_2}
A.~Le~Glaz, Y.~Haralambous, D-H. Kim-Dufor, P.~Lenca, R.~Biliot, T.~C. Ryan, J.~Marsh, J.~DeVylder, M.~Walter, S.~Berrouiguet, and C.~Lemey.
\newblock Machine learning and natural language processing in mental health: Systematic review.
\newblock {\em Journal of Medical Internet Research}, 23(5), 2021.

\bibitem{intro_17_1}
F.~Yu, C.~Wei, P.~Deng, T.~Peng, and X.~Hu.
\newblock Deep exploration of random forest model boosts the interpretability of machine learning studies of complicated immune responses and lung burden of nanoparticles.
\newblock {\em Science advances}, 7(22):eabf4130, 2021.

\bibitem{intro_17_2}
M.~Pavlovi{\'c}, L.~Scheffer, K.~Motwani, C.~Kanduri, R.~Kompova, N.~Vazov, K.~Waagan, F.~L.~M. Bernal, A.~A. Costa, B.~Corrie, et~al.
\newblock The immuneml ecosystem for machine learning analysis of adaptive immune receptor repertoires.
\newblock {\em Nature Machine Intelligence}, 3(11):936--944, 2021.

\bibitem{Antoni2006TheIO}
Michael~H Antoni, Susan~K. Lutgendorf, Steve~W. Cole, Firdaus~S. Dhabhar, Sandra~E. Sephton, Paige~Green McDonald, Michael~Edward Stefanek, and Anil~K. Sood.
\newblock The influence of bio-behavioural factors on tumour biology: pathways and mechanisms.
\newblock {\em Nature Reviews Cancer}, 6:240--248, 2006.

\bibitem{Zhang2025BurdenAR}
Tao Zhang, Shuai Wang, Dongming Li, Yifei Wang, and Xueyuan Cao.
\newblock Burden and risk factors of colorectal cancer in europe from 1990 to 2021.
\newblock {\em European journal of cancer prevention : the official journal of the European Cancer Prevention Organisation}, 2025.

\bibitem{Cui2020CancerAS}
Bai Cui, Fei Peng, Jinxin Lu, Bin He, Qitong Su, Huandong Luo, Ziqian Deng, Tonghui Jiang, Keyu Su, Yanping Huang, Zaheer~Ud Din, Eric Wing-Fai Lam, Keith~W. Kelley, and Quentin Liu.
\newblock Cancer and stress: Nextgen strategies.
\newblock {\em Brain, Behavior, and Immunity}, 93:368--383, 2020.

\bibitem{Kruk2019PsychologicalSA}
J.~Kruk, B.~H. Aboul-Enein, J.~Bernstein, and M.~Gronostaj.
\newblock Psychological stress and cellular aging in cancer: A meta-analysis.
\newblock {\em Oxidative Medicine and Cellular Longevity}, 2019, 2019.

\bibitem{Dai2020ChronicSP}
S.~Dai, Y.~Mo, Y.~Wang, B.~Xiang, Q.~Liao, M.~Zhou, X.~Li, Y.~Li, W.~Xiong, G-Y. Li, C.~Guo, and Z.~Zeng.
\newblock Chronic stress promotes cancer development.
\newblock {\em Frontiers in Oncology}, 10, 2020.

\bibitem{Chiriac2017PsychologicalSA}
V.~Chiriac, A.~Baban, and D.~L. Dumitrascu.
\newblock Psychological stress and breast cancer incidence: a systematic review.
\newblock {\em Clujul Medical}, 91:18 -- 26, 2017.

\bibitem{Bahri2019TheRB}
Narjes Bahri, Tahereh~Fathi Najafi, Fatemeh~Homaei Shandiz, Hamid~Reza Tohidinik, and Abdoljavad Khajavi.
\newblock The relation between stressful life events and breast cancer: a systematic review and meta-analysis of cohort studies.
\newblock {\em Breast Cancer Research and Treatment}, 176:53--61, 2019.

\bibitem{Lempesis2023RoleOS}
I.~G. Lempesis, V.~E. Georgakopoulou, P.~Papalexis, G.~P. Chrousos, and D.~A. Spandidos.
\newblock Role of stress in the pathogenesis of cancer (review).
\newblock {\em International Journal of Oncology}, 63, 2023.

\bibitem{Heikkil2013WorkSA}
Katriina Heikkil{\"a}, Solja~T. Nyberg, T{\"o}res Theorell, Eleonor~I. Fransson, Lars Alfredsson, Jakob~Bue Bjorner, S{\'e}bastien Bonenfant, Marianne Borritz, Kim Bouillon, Hermann Burr, Nico Dragano, Goedele~A. Geuskens, Marcel Goldberg, Mark Hamer, Wendela~E. Hooftman, Irene L.~D. Houtman, Matti Joensuu, Anders Knutsson, Markku Koskenvuo, Aki Koskinen, Anne~M Kouvonen, Ida E~H Madsen, Linda L.~Magnusson Hanson, Michael Marmot, Martin~Lindhardt Nielsen, Maria Nordin, Tuula Oksanen, Jaana Pentti, Paula~M. Salo, Reiner Rugulies, Andrew Steptoe, Sakari~B Suominen, Jussi Vahtera, Marianna Virtanen, Ari V{\"a}{\"a}n{\"a}nen, Peter J.~M. Westerholm, Hugo Westerlund, Marie Zins, Jane~E. Ferrie, Archana Singh‐Manoux, G.~David Batty, and Mika Kivim{\"a}ki.
\newblock Work stress and risk of cancer: meta-analysis of 5700 incident cancer events in 116000 european men and women.
\newblock {\em The BMJ}, 346, 2013.

\bibitem{CortsIbez2022TheVO}
Francisco~O. Cort{\'e}s-Ib{\'a}{\~n}ez, Bram van Pinxteren, Anna Sijtsma, Annette~H. Bruggink, Grigory Sidorenkov, Bert van~der Vegt, and Geertruida~H. de~Bock.
\newblock The validity of self-reported cancer in a population-based cohort compared to that in formally registered sources.
\newblock {\em Cancer epidemiology}, 81:102268, 2022.

\bibitem{Avgerinos2019ObesityAC}
Konstantinos~Ioannis Avgerinos, Nikolaos~K. Spyrou, Christos~Socrates Mantzoros, and Maria Dalamaga.
\newblock Obesity and cancer risk: Emerging biological mechanisms and perspectives.
\newblock {\em Metabolism: clinical and experimental}, 92:121--135, 2019.

\bibitem{Pham2024ChronicSR}
An~Thanh Pham, Boukje A~C van Dijk, Eline~S van~der Valk, Bert van~der Vegt, Elisabeth F~C van Rossum, and Geertruida~H. de~Bock.
\newblock Chronic stress related to cancer incidence, including the role of metabolic syndrome components.
\newblock {\em Cancers}, 16, 2024.

\bibitem{Alotiby2024ImmunologyOS}
A.~Alotiby.
\newblock Immunology of stress: A review article.
\newblock {\em Journal of Clinical Medicine}, 13, 2024.

\bibitem{Nakata2012PsychosocialJS}
A.~Nakata.
\newblock Psychosocial job stress and immunity: a systematic review.
\newblock {\em Methods in molecular biology}, 934:39--75, 2012.

\bibitem{segerstrom2004psychological}
S.~C. Segerstrom and G.~E. Miller.
\newblock Psychological stress and the human immune system: a meta-analytic study of 30 years of inquiry.
\newblock {\em Psychological bulletin}, 130(4):601, 2004.

\bibitem{Dhabhar2013PsychologicalSA}
F.~S. Dhabhar.
\newblock Psychological stress and immunoprotection versus immunopathology in the skin.
\newblock {\em Clinics in dermatology}, 31 1:18--30, 2013.

\bibitem{intro_8_1}
B.~Mueller, A.~Figueroa, and J.~Robinson-Papp.
\newblock Structural and functional connections between the autonomic nervous system, hypothalamic--pituitary--adrenal axis, and the immune system: a context and time dependent stress response network.
\newblock {\em Neurological sciences}, 43(2):951--960, 2022.

\bibitem{intro_8_2}
P.~H. Black.
\newblock Central nervous system-immune system interactions: psychoneuroendocrinology of stress and its immune consequences.
\newblock {\em Antimicrobial agents and chemotherapy}, 38(1):1--6, 1994.

\bibitem{intro_9_1}
F.~S. Dhabhar.
\newblock Effects of stress on immune function: the good, the bad, and the beautiful.
\newblock {\em Immunologic research}, 58:193--210, 2014.

\bibitem{intro_9_2}
J.~J. Kamma, C.~Giannopoulou, V.~G.~S. Vasdekis, and A.~Mombelli.
\newblock Cytokine profile in gingival crevicular fluid of aggressive periodontitis: influence of smoking and stress.
\newblock {\em Journal of clinical periodontology}, 31(10):894--902, 2004.

\bibitem{Gouin2008ImmuneDA}
J-P. Gouin, L.~Hantsoo, and J.~K. Kiecolt-Glaser.
\newblock Immune dysregulation and chronic stress among older adults: A review.
\newblock {\em Neuroimmunomodulation}, 15:251 -- 259, 2008.

\bibitem{Hong2021ChronicSE}
H.~Hongm, M.~Ji, and D.~Lai.
\newblock Chronic stress effects on tumor: Pathway and mechanism.
\newblock {\em Frontiers in Oncology}, 11, 2021.

\bibitem{ColonEchevarria2019NeuroendocrineRO}
C.~B. Colon-Echevarria, R.~Lamboy-Caraballo, A.~N. Aquino-Acevedo, and G.~N. Armaiz-Pena.
\newblock Neuroendocrine regulation of tumor-associated immune cells.
\newblock {\em Frontiers in Oncology}, 9, 2019.

\bibitem{Kusnecov2002StressorinducedMO}
A.~W. Kusnecov and A.~Rossi-George.
\newblock Stressor-induced modulation of immune function: a review of acute, chronic effects in animals.
\newblock {\em Acta Neuropsychiatrica}, 14:279 -- 291, 2002.

\bibitem{Silverman2012GlucocorticoidRO}
M.~N. Silverman and E.~M. Sternberg.
\newblock Glucocorticoid regulation of inflammation and its functional correlates: from hpa axis to glucocorticoid receptor dysfunction.
\newblock {\em Annals of the New York Academy of Sciences}, 1261, 2012.

\bibitem{intro_13_1}
F.~S. Dhabhar.
\newblock Enhancing versus suppressive effects of stress on immune function: implications for immunoprotection versus immunopathology.
\newblock {\em Allergy, Asthma \& Clinical Immunology}, 4(1):2, 2008.

\bibitem{intro_13_2}
J~Szelenyi and E.~S. Vizi.
\newblock The catecholamine--cytokine balance: interaction between the brain and the immune system.
\newblock {\em Annals of the New York Academy of Sciences}, 1113(1):311--324, 2007.

\bibitem{intro_10}
J.~O’Donnell-Tormey and M.~Tontonoz.
\newblock Cancer and the immune system: the vital connection.
\newblock {\em Cancer Research Institute}, pages 7--8, 2016.

\bibitem{intro_12_1}
L.~Fiorentino and S.~Ancoli-Israel.
\newblock Sleep dysfunction in patients with cancer.
\newblock {\em Current treatment options in neurology}, 9(5):337--346, 2007.

\bibitem{intro_12_2}
J.~A. Paice.
\newblock Chronic treatment-related pain in cancer survivors.
\newblock {\em Pain}, 152(3):S84--S89, 2011.

\bibitem{intro_12_3}
J.~E. Smith, J.~Richardson, C.~Hoffman, and K.~Pilkington.
\newblock Mindfulness-based stress reduction as supportive therapy in cancer care: systematic review.
\newblock {\em Journal of advanced nursing}, 52(3):315--327, 2005.

\bibitem{Cohen2001PsychologicalSA}
S.~Cohen, G.~E. Miller, and G.~S. Rabin.
\newblock Psychological stress and antibody response to immunization: A critical review of the human literature.
\newblock {\em Psychosomatic Medicine}, 63:7--18, 2001.

\bibitem{MorenoSmith2010ImpactOS}
Myrthala Moreno-Smith, Susan~K. Lutgendorf, and Anil~K. Sood.
\newblock Impact of stress on cancer metastasis.
\newblock {\em Future oncology}, 6 12:1863--81, 2010.

\bibitem{Rolln2022CurrentKO}
M.~P. Roll{\'a}n, R.~N. Cabrera, and R.~A. Schwartz.
\newblock Current knowledge of immunosuppression as a risk factor for skin cancer development.
\newblock {\em Critical reviews in oncology/hematology}, page 103754, 2022.

\bibitem{doss2016changing}
Mohan Doss.
\newblock Changing the paradigm of cancer screening, prevention, and treatment.
\newblock {\em Dose-Response}, 14(4):1559325816680539, 2016.

\bibitem{Zhang2020ChronicSI}
L.~Zhang, J.~Pan, W.~Chen, J.~Jiang, and J.~Huang.
\newblock Chronic stress-induced immune dysregulation in cancer: implications for initiation, progression, metastasis, and treatment.
\newblock {\em American journal of cancer research}, 10 5:1294--1307, 2020.

\bibitem{Liu2014TheRO}
Yunxin Liu, Xianjun Fang, Jie Yuan, Zongxing Sun, Chuanhua Li, Rong-Gang Li, Li~Li, Chao Zhu, Rong Wan, Rui Guo, Lai Jin, and Shengnan Li.
\newblock The role of corticotropin-releasing hormone receptor 1 in the development of colitis-associated cancer in mouse model.
\newblock {\em Endocrine-related cancer}, 21 4:639--51, 2014.

\bibitem{Troche2014SystemicGU}
J.~R. Troche, L.~M. Ferrucci, B.~Cartmel, D.~J. Leffell, A.~E. Bale, and S.~T. Mayne.
\newblock Systemic glucocorticoid use and early-onset basal cell carcinoma.
\newblock {\em Annals of epidemiology}, 24 8:625--7, 2014.

\bibitem{Kareva2020ImmuneSI}
I.~Kareva.
\newblock Immune suppression in pregnancy and cancer: Parallels and insights.
\newblock {\em Translational Oncology}, 13, 2020.

\bibitem{Stewart2011ImprovingCI}
T.~J. Stewart and M.~J. Smyth.
\newblock Improving cancer immunotherapy by targeting tumor-induced immune suppression.
\newblock {\em Cancer and Metastasis Reviews}, 30:125--140, 2011.

\bibitem{Coussens2013NeutralizingTC}
L.~M. Coussens, L.~Zitvogel, and A.~K. Palucka.
\newblock Neutralizing tumor-promoting chronic inflammation: A magic bullet?
\newblock {\em Science}, 339:286 -- 291, 2013.

\bibitem{Bajpai2020CancerIF}
J.~Bajpai.
\newblock Cancer immunotherapy for immunocompromised patients: An often ignored, yet vital puzzle.
\newblock {\em Journal of Immunotherapy and Precision Oncology}, 3:1 -- 2, 2020.

\bibitem{qu_1}
N.~Harder, L.~Figueroa, R.~M. Gillum, D.~Hangartner, D.~D. Laitin, and J.~Hainmueller.
\newblock Multidimensional measure of immigrant integration.
\newblock {\em Proceedings of the National Academy of Sciences}, 115(45):11483--11488, 2018.

\bibitem{qu_2}
E.~Savchenko, A.~Rosenfeld, and S.~Bunimovich-Mendrazitsky.
\newblock A mathematical framework of sms reminder campaigns for pre-and post-diagnosis check-ups using socio-demographics: An in-silco investigation into breast cancer.
\newblock {\em Socio-Economic Planning Sciences}, 95:102047, 2024.

\bibitem{qu_3}
A.~Yaniv-Rosenfeld, E.~Savchenko, A.~Elalouf, and U.~Nitzan.
\newblock Socio-demographic predictors of the time interval between successive hospitalizations among patients with borderline personality disorder.
\newblock {\em Journal of Mental Health}, pages 1--7, 2024.

\bibitem{qu_4}
D.~Orentlicher.
\newblock Healthcare, health, and income.
\newblock {\em The Journal of Law, Medicine \& Ethics}, 46(3):567--572, 2018.

\bibitem{qu_5}
N.~Struch, Y.~Shereshevsky, A.~Baidani-Auerbach, M.~Lachman, N.~Sagiv, T.~Zehavi, and I.~Levav.
\newblock Attitudes, knowledge and preferences of the israeli public regarding mental health services.
\newblock {\em Israel Journal of Psychiatry and Related Sciences}, 45(2):129, 2008.

\bibitem{s_second_1}
P.~Decoufle.
\newblock {\em A retrospective survey of cancer in relation to occupation}.
\newblock Number 77-178. US Department of Health, Education, and Welfare, Public Health Service~…, 1977.

\bibitem{s_second_2}
X.~Chen and L.~L. Siu.
\newblock Impact of the media and the internet on oncology: survey of cancer patients and oncologists in canada.
\newblock {\em Journal of Clinical Oncology}, 19(23):4291--4297, 2001.

\bibitem{s_third_1}
P.~L. Mai, A.~O. Garceau, B.~I. Graubard, M.~Dunn, T.~S. McNeel, L.~Gonsalves, M.~H. Gail, M.~H. Greene, G.~B. Willis, and L.~Wideroff.
\newblock Confirmation of family cancer history reported in a population-based survey.
\newblock {\em Journal of the National Cancer Institute}, 103(10):788--797, 2011.

\bibitem{s_third_2}
H.~T. Lynch, H.~A. Guirgis, P.~M. Lynch, J.~F. Lynch, and R.~E. Harris.
\newblock Familial cancer syndromes: a survey.
\newblock {\em Cancer}, 39(S4):1867--1881, 1977.

\bibitem{s_third_3}
C.~C. Teerlink, F.~S. Albright, L.~Lins, and L.~A. Cannon-Albright.
\newblock A comprehensive survey of cancer risks in extended families.
\newblock {\em Genetics in Medicine}, 14(1):107--114, 2012.

\bibitem{s_last_1}
A.~Caraceni and R.~K. Portenoy.
\newblock An international survey of cancer pain characteristics and syndromes.
\newblock {\em Pain}, 82(3):263--274, 1999.

\bibitem{s_last_2}
B.~Vanaelst, N.~Michels, E.~Clays, D.~Herrmann, I.~Huybrechts, I.~Sioen, K.~Vyncke, and S.~De~Henauw.
\newblock The association between childhood stress and body composition, and the role of stress-related lifestyle factors—cross-sectional findings from the baseline chibs survey.
\newblock {\em International journal of behavioral medicine}, 21:292--301, 2014.

\bibitem{s_last_3}
R.~S. Nastaskin and A.~J. Fiocco.
\newblock A survey of diet self-efficacy and food intake in students with high and low perceived stress.
\newblock {\em Nutrition journal}, 14:1--8, 2015.

\bibitem{s_last_4}
B.~S. McEwen.
\newblock Neurobiological and systemic effects of chronic stress.
\newblock {\em Chronic stress}, 1:2470547017692328, 2017.

\bibitem{joshi2015likert}
A.~Joshi, S.~Kale, S.~Chandel, and D.~K. Pal.
\newblock Likert scale: Explored and explained.
\newblock {\em British journal of applied science \& technology}, 7(4):396--403, 2015.

\bibitem{vasantha2016online}
N.~N. S.~H. Vasantha~Raju and N.~S. Harinarayana.
\newblock Online survey tools: A case study of google forms.
\newblock In {\em National conference on scientific, computational \& information research trends in engineering, GSSS-IETW, Mysore}, 2016.

\bibitem{python}
K.~R. Srinath.
\newblock Python – the fastest growing programming language.
\newblock {\em International Research Journal of Engineering and Technology}, 4(12), 2017.

\bibitem{pearson}
D.~Freedman, R.~Pisani, and R.~Purves.
\newblock Statistics (international student edition).
\newblock {\em Pisani, R. Purves, 4th edn. WW Norton \& Company, New York}, 2007.

\bibitem{anova}
E.~R. Girden.
\newblock {\em ANOVA: Repeated measures}.
\newblock Number~84. Sage, 1992.

\bibitem{t_test}
D.~Kalpi{\'{c}}, N.~Hlupi{\'{c}}, and M.~Lovri{\'{c}}.
\newblock {\em Student's t-Tests}, pages 1559--1563.
\newblock 2011.

\bibitem{draper1998applied}
N.~R. Draper and H.~Smith.
\newblock {\em Applied regression analysis}, volume 326.
\newblock John Wiley \& Sons, 1998.

\bibitem{lg_1}
L.~Shami and T.~Lazebnik.
\newblock Implementing machine learning methods in estimating the size of the non-observed economy.
\newblock {\em Computational Economics}, 2023.

\bibitem{dt_1}
G.~Bonner.
\newblock Decision making for health care professionals: use of decision trees within the community mental health setting.
\newblock {\em Journal of Advanced Nursing}, 35:349--356, 2001.

\bibitem{xgboost}
A.~Shmuel and E.~Heifetz.
\newblock Developing novel machine-learning-based fire weather indices.
\newblock {\em Machine Learning: Science and Technology}, 4(1):015029, 2023.

\bibitem{knn}
S.~Dhanabal and S.~Chandramathi.
\newblock A review of various k-nearest neighbor query processing techniques.
\newblock {\em International Journal of Computer Applications}, 31(7):14--22, 2011.

\bibitem{dt_2}
S.~Letourneau and L.~Jensen.
\newblock Impact of a decision tree on chronic wound care.
\newblock {\em Wound Ostomy Continence Nurs}, 25:240--247, 1998.

\bibitem{dt_3}
A.~S. Abdullah, S.~Selvakumar, P.~Karthikeyan, and M.~Venkatesh.
\newblock Comparing the efficacy of decision tree and its variants using medical data.
\newblock {\em Indian Journal of Science and Technology}, 10:1--8, 2017.

\bibitem{shmuel2024comprehensive}
A.~Shmuel, O.~Glickman, and T.~Lazebnik.
\newblock A comprehensive benchmark of machine and deep learning across diverse tabular datasets.
\newblock {\em arXiv}, 2024.

\bibitem{carbonell1983overview}
J.~G. Carbonell, R.~S. Michalski, and T.~M. Mitchell.
\newblock An overview of machine learning.
\newblock {\em Machine learning}, pages 3--23, 1983.

\bibitem{wong2019reliable}
T-T. Wong and P-Y. Yeh.
\newblock Reliable accuracy estimates from k-fold cross validation.
\newblock {\em IEEE Transactions on Knowledge and Data Engineering}, 32(8):1586--1594, 2019.

\bibitem{jung2018multiple}
Y.~Jung.
\newblock Multiple predicting k-fold cross-validation for model selection.
\newblock {\em Journal of nonparametric statistics}, 30(1):197--215, 2018.

\bibitem{ponv_6}
F.~Doshi-Velez and R.~H. Perlis.
\newblock Evaluating machine learning articles.
\newblock {\em JAMA}, 322(18):1777--1779, 2019.

\bibitem{ponv_7}
K.~Dowsland and J.~Thompson.
\newblock Solving a nurse scheduling problem with knapsacks, networks and tabu search.
\newblock {\em J Oper Res Soc}, 51:825--833, 2000.

\bibitem{ponv_9}
S.~L. Goh, S.~N. Sze, N.~R. Sabar, S.~Abdullah, and G.~Kendall.
\newblock A 2-stage approach for the nurse rostering problem.
\newblock {\em IEEE Access}, 10:69591--69604, 2022.

\bibitem{ponv_10}
M.~D. Goodman, K.~A. Dowsland, and J.~M. Thompson.
\newblock A grasp-knapsack hybrid for a nurse-scheduling problem.
\newblock {\em Journal of Heuristics}, 15(3):351--379, 2007.

\bibitem{ponv_24}
M.~Rajeswari, J.~Amudhavel, S.~Pothula, and P.~Dhavachelvan.
\newblock Directed bee colony optimization algorithm to solve the nurse rostering problem.
\newblock {\em Computational Intelligence and Neuroscience}, 2017.

\bibitem{alibrahim2021hyperparameter}
H.~Alibrahim and S.~A. Ludwig.
\newblock Hyperparameter optimization: Comparing genetic algorithm against grid search and bayesian optimization.
\newblock In {\em 2021 IEEE congress on evolutionary computation (CEC)}, pages 1551--1559. IEEE, 2021.

\bibitem{mantovani2018empirical}
R.~G. Mantovani, T.~Horv{\'a}th, R.~Cerri, S.~B. Junior, J.~Vanschoren, and A.~C.~D. de~Carvalho.
\newblock An empirical study on hyperparameter tuning of decision trees.
\newblock {\em arXiv}, 2018.

\bibitem{glebov2023predicting}
M.~Glebov, T.~Lazebnik, B.~Orkin, H.~Berkenstadt, and S.~Bunimovich-Mendrazitsky.
\newblock Predicting postoperative nausea and vomiting using machine learning: A model development and validation study.
\newblock {\em arXiv}, 2023.

\bibitem{rizki2024optimization}
M.~Rizki, A.~Hermawan, and D.~Avianto.
\newblock Optimization of hyperparameter k in k-nearest neighbor using particle swarm optimization.
\newblock {\em JUITA: Jurnal Informatika}, 12(1):71--79, 2024.

\bibitem{Benchmarking_AutoML}
E.~A. Neverov, I.~I. Viksnin, and S.~S. Chuprov.
\newblock The research of automl methods in the task of wave data classification.
\newblock In {\em 2023 XXVI International Conference on Soft Computing and Measurements (SCM)}, pages 156--158, 2023.

\bibitem{Benchmarking_AutoML2}
P.~H. Ribeiro, P.~Orzechowski, J.~Wagenaar, and J.~H. Moore.
\newblock Benchmarking automl algorithms on a collection of synthetic classification problems.
\newblock {\em CoRR}, abs/2212.02704, 2022.

\bibitem{fi_1}
A.~Zien, N.~Kramer, S.~Sonnenburg, and G.~Ratsch.
\newblock The feature importance ranking measure.
\newblock In W.~Buntine, M.~Grobelnik, D.~Mladenic, and J.~Shawe-Taylor, editors, {\em Machine Learning and Knowledge Discovery in Databases}, pages 694--709. Springer Berlin Heidelberg, 2009.

\bibitem{fi_2}
G.~Casalicchio, C.~Molnar, and B.~Bischl.
\newblock Visualizing the feature importance for black box models.
\newblock In M.~Berlingerio, F.~Bonchi, T.~Gartner, N.~Hurley, and G.~Ifrim, editors, {\em Machine Learning and Knowledge Discovery in Databases}, pages 655--670. Springer International Publishing, 2019.

\bibitem{fi_3}
A.~Altmann, L.~Tolosi, O.~Sander, and T.~Lengauer.
\newblock Permutation importance: a corrected feature importance measure.
\newblock {\em Bioinformatics}, 26(10):1340--1347, 2010.

\bibitem{Mayrhofer_Filzmoser_2023}
M.~Mayrhofer and P.~Filzmoser.
\newblock Multivariate outlier explanations using shapley values and mahalanobis distances.
\newblock {\em Econometrics and Statistics}, 2023.

\bibitem{meng2020makes}
Y.~Meng, N.~Yang, Z.~Qian, and G.~Zhang.
\newblock What makes an online review more helpful: an interpretation framework using xgboost and shap values.
\newblock {\em Journal of Theoretical and Applied Electronic Commerce Research}, 16(3):466--490, 2020.

\bibitem{imbens2010rubin}
G.~W. Imbens and D.~B. Rubin.
\newblock Rubin causal model.
\newblock In {\em Microeconometrics}, pages 229--241. Springer, 2010.

\bibitem{benedetto2018statistical}
U.~Benedetto, S.~J. Head, G.~D. Angelini, and E.~H. Blackstone.
\newblock Statistical primer: propensity score matching and its alternatives.
\newblock {\em European Journal of Cardio-Thoracic Surgery}, 53(6):1112--1117, 2018.

\bibitem{hanusz2014simulation}
Z.~Hanusz and J.~Tarasi{\'N}ska.
\newblock Simulation study on improved shapiro--wilk tests for normality.
\newblock {\em Communications in Statistics-Simulation and Computation}, 43(9):2093--2105, 2014.

\bibitem{Bauer2009TheRO}
M.~E. Bauer, C.~M.~M. Jeckel, and C.~Luz.
\newblock The role of stress factors during aging of the immune system.
\newblock {\em Annals of the New York Academy of Sciences}, 1153, 2009.

\end{thebibliography}
\bibliographystyle{unsrt}

\newpage

\appendix
\section*{Appendix}
\section{Questionnaire}\label{app:survey}
The online study consisted of the following questions and possible answers as closed-form questions.
\begin{itemize}
    \item \say{Country}: Open-ended.
    \item \say{Gender}: Male, Female, Prefer not to say.
    \item \say{Age}: 18-120.
    \item \say{What is your material status?}: Single, in a relationship, married, devoured, widow/er. 
    \item \say{How many children do you have?}: 0, 1, 2, 3, 4, 5+.
    \item \say{Education}: Did not complete high-school, High-School, Bachelor's, Master's degree, PhD/M.D.
    \item \say{Household Monthly Income (before taxes and deductions) from all sources}: up to 2150 USD monthly, 2,150–2,700 USD monthly, 2,700–3,250 USD monthly, 3,250–5,400 USD monthly, 5,400–8,100 USD monthly, between 8,100–10,800 USD monthly, 10,800 USD monthly, would like not to answer.
    \item \say{How many people live with you in the same house?}: 1-15.

    \item \say{Have you ever been diagnosed with cancer?}: Yes, No.
    \item \say{If yes, what type of cancer were you diagnosed with?}: Open-ended.
    \item \say{If yes, when were you diagnosed?}: Less than 1 year ago, 1-3 years ago, 3-5 years ago, Over 5 years ago.
    \item \say{Are you currently undergoing cancer treatment?}: Yes, No.

    \item \say{Have any of your immediate family members been diagnosed with cancer?}: Yes, No.
    \item \say{If yes, how many family members have been diagnosed with cancer?}: 1, 2, 3, 4, 5+.
    \item \say{If yes, What is the relationship to the closest diagnosed family members?}: Parent, Sibling, Child, Other (open-question).
    \item \say{If yes, what type of cancer was your closest family member diagnosed with?}: Open-ended.

    \item \say{How often do you feel stressed in your daily life?}: Never, Rarely, Sometimes, Often, All the time.
    \item \say{Do you believe stress affects your physical health?}: Strongly agree, Agree, Neutral, Disagree, Strongly disagree.
    \item \say{How well do you cope with stress?}: Very well, Well, Average, Poorly, Very poorly.
    \item \say{In the past month, how often have you experienced difficulty sleeping?}: Never, Rarely, Sometimes, Often, All the time.
    \item \say{In the past month, how often have you experienced loss of appetite?}: Never, Rarely, Sometimes, Often, All the time.
    \item \say{In the past month, how often have you experienced fatigue?}: Never, Rarely, Sometimes, Often, All the time.
    \item \say{In the past month, how often have you experienced difficulty concentrating?}: Never, Rarely, Sometimes, Often, All the time.
    \item \say{How much do you feel that stress has impacted your mental health in the past year?}: Not at all, A little, Moderately, Significantly, Extremely.
\end{itemize}

\end{document}